\documentclass[aps,pre,reprint,superscriptaddress]{revtex4-2}
\usepackage{iftex}
\ifpTeX
  \PassOptionsToPackage{dvipdfmx}{color}
  \PassOptionsToPackage{dvipdfmx}{xcolor}
  \PassOptionsToPackage{dvipdfmx}{graphicx}
  \PassOptionsToPackage{dvipdfmx}{hyperref}
\fi
\usepackage{bm}
\usepackage{physics}
\usepackage{anyfontsize}
\usepackage{amsmath}
\usepackage{amsthm}
\usepackage{color}
\usepackage{comment} 
\usepackage{array}
\usepackage{amssymb}

\usepackage{mathtools}
\usepackage{newtxtext}
\usepackage{graphicx}
\usepackage[colorlinks=true,allcolors=blue]{hyperref}
\usepackage{etoolbox}
\makeatletter
\patchcmd{\@bibdataout@aps}{author="08"}{author="48"}{}{}
\patchcmd{\@bibdataout@aps}{author="08"}{author="48"}{}{}
\makeatother
\newcommand{\red}[1]{\textcolor{red}{#1}}

\newcommand{\f}{\frac}

\begin{document}
\title{Thermodynamic Speed Limits in Isolated Quantum Systems}

\author{Rikuya Kobashi}
\email{kobashi.rikuya.84s@st.kyoto-u.ac.jp}
\affiliation{Department of Physics, Kyoto University, Kyoto 606-8502, Japan}

\begin{abstract}
{
We derive thermodynamic speed limits for isolated quantum systems undergoing finite-time unitary driving. The key ingredient is a set of generalized second-law inequalities: for the Gibbs entropy and for several thermodynamically motivated observational entropies, the entropy production is bounded from below by a function of the Vu--Saito quantum Wasserstein distance between the initial and final coarse-grained states. These inequalities yield lower bounds on the operation time in terms of the average entropy-production rate. We apply the framework to system--bath coarse-graining, local-energy coarse-graining, and diagonal entropy, and illustrate the resulting bounds with numerical and analytically solvable examples. Our results provide a thermodynamic characterization of finite-time state transformations in isolated quantum systems and clarify how coarse-grained entropy production constrains macroscopic reachability.}
\end{abstract}

\maketitle

\section{Introduction}\label{sec:intro}
Is it possible, in principle, to move through space or to make computers perform calculations at arbitrarily high speed? Prior studies indicate that fundamental physical laws constrain not only energetic and entropic resources but also the rate at which physical states—and hence information processing—can be transformed. In isolated quantum systems, this constraint is formalized as the quantum speed limit (QSL): the minimum time required to implement a given evolution is bounded from below by the energy fluctuations \cite{Mandelstam1991} or the mean energy \cite{MargolusLevitin1998}, together with a geometric distance between the initial and final states \cite{JonesKok2010, Deffner2017}. Such speed-limit inequalities are important not only from a fundamental perspective but also for a wide range of tasks based on isolated quantum systems—such as state transfer, optimal control, and related tasks—where they set intrinsic bounds on performance \cite{Yung2006, Caneva2009, Seth2000, DeffnerLuts2010}.

{Speed limits are not unique to isolated quantum systems. For open systems, whose dynamics is constrained by thermodynamics, it is natural to ask how speed limits are related to thermodynamic quantities. In the field of stochastic thermodynamics, the rate of change of the state distribution is constrained by trade-offs involving entropy production and dynamical activity \cite{Shiraishifunosaito2018, Ito2018, ItoDechant2020, Tuanetal2020, YoshimuraIto2021}. These bounds are now regarded as fundamental constraints in nonequilibrium thermodynamics. Similar speed-limit relations have also been reported for Lindblad-type open quantum systems \cite{Funo2019}.}

Recent progress in stochastic thermodynamics has been further propelled by incorporating optimal transport theory \cite{Aurelletal2011}. Optimal transport theory is a mathematical framework that has been developed to construct optimal protocols for transforming one probability distribution into another \cite{Villani2009}. Many tasks realized using physical systems can be viewed precisely as processes that transform an initial distribution into a desired target distribution. It is therefore not surprising that insights from physics concerning such tasks are closely related to optimal transport theory. In particular, it is known that, in stochastic thermodynamic systems, the minimum entropy production required to perform a given task is intimately connected to the $L_2$-Wasserstein distance familiar in the overdamped Langevin dynamics \cite{BenamouBrenier2000}. Moreover, this interplay between thermodynamics and optimal transport has also been developed for the Markov jump process as well as  Lindblad-type open quantum systems \cite{VuSaito2023}. By defining a quantum Wasserstein distance as a quantum extension of the classical Wasserstein metric, previous work derived a Benamou--Brenier-type identity and established new speed-limit bounds.

Understanding how phenomena that appear irreversible on macroscopic scales can be reconciled with the reversible dynamics of isolated quantum systems has long been a fundamental problem.
With recent advances in quantum simulators such as ultracold atoms and trapped-ion platforms, the problem of thermalization and equilibration in isolated quantum systems has seen rapid progress on both theoretical and experimental fronts \cite{Tasaki1998,Short2011,Trotzky2012, Cramer2010,ShiraishiTasaki2024, Kaufman2016}. In parallel, substantial effort has been devoted to examining whether an isolated quantum system, once equilibrated, is consistent with macroscopic thermodynamic laws. In particular, a long-running line of work has focused on microscopic formulations of thermodynamic quantities. Within this approach, thermodynamic behavior, including the second law, is derived directly from the unitary dynamics of isolated quantum systems in a variety of settings\cite{Tasaki2016, Hokkyo2025, Meier2025, Brandao2015}. A basic obstacle is that under strictly unitary time evolution the von Neumann entropy is conserved, and thus cannot by itself capture irreversibility. To address this, alternative entropy notions have been proposed, notably, observational entropy \cite{SafranekDeutschAguirre2019}—including the diagonal entropy \cite{Ikeda2015,Polkovnikov2011}—and the Gibbs entropy \cite{Gibbs1902,tasaki2000}. These entropies satisfy thermodynamic differential relations when the system is in equilibrium and can increase under driving, thereby providing entropy measures that are compatible with the second law for isolated quantum dynamics.

As reviewed above, thermal behavior can emerge even in isolated systems. However, in contrast to open-system, finite-time thermodynamic constraints for isolated dynamics---e.g. thermodynamic speed limits---still appear to be far from fully understood. At the same time, to the best of our knowledge, the connection between nonequilibrium phenomena in isolated systems and optimal transport theory remains largely unexplored. In light of these considerations, we build on the recent understanding of thermalization in isolated quantum systems and on microscopic derivations of the second law. We then develop a more quantitative characterization of entropy production, based on thermodynamically meaningful entropies and quantified using a quantum Wasserstein distance.

{
The main contributions of this work are threefold. First, we derive generalized second-law inequalities for isolated quantum systems, in which entropy production associated with the Gibbs entropy and observational entropy is bounded from below by the Vu--Saito quantum Wasserstein distance. Second, we apply these inequalities to thermodynamically motivated coarse-grainings, including system--bath, local-energy, and energy-basis coarse-grainings. Third, by dividing the entropy-production bounds by the average entropy-production rate, we obtain thermodynamic speed limits for finite-time unitary protocols.

The remainder of this paper is organized as follows. In Sec.~\ref{sec:model}, we introduce the setup. In Sec.~\ref{sec:ent-wass}, we define the Gibbs entropy, observational entropy, and the Vu--Saito quantum Wasserstein distance. In Sec.~\ref{sec:main_results}, we present the generalized second laws and derive the corresponding speed limits. In Sec.~\ref{sec:APP}, we discuss applications and numerical demonstrations. In Sec.~\ref{sec:Deriv}, we provide the derivations of the main results. Section~\ref{sec:remarks} is devoted to conclusions and outlook.
}

\section{Setup}\label{sec:model}

{
We consider an isolated quantum system on a $d$-dimensional Hilbert space $\mathcal H$. Its state is represented by a density operator $\rho$, satisfying $\rho\geq 0$ and $\tr[\rho]=1$. The dynamics are generated by a generally time-dependent Hamiltonian $H(\lambda_t)$, where $\lambda_t$ denotes an external control parameter. We write the spectral decomposition as
\begin{align}
H(\lambda)=\sum_{i=1}^{n}E_i(\lambda)\hat P_i^{(E)}(\lambda),
\end{align}
where $E_i(\lambda)$ is the $i$-th energy eigenvalue in increasing order, and $\hat P_i^{(E)}(\lambda)$ is the corresponding spectral projector.

We use projective measurements as coarse-grainings of the Hilbert space. A projection-valued measure (PVM) $\mathcal C=\{\hat P_x\}_{x\in\mathcal X}$ satisfies $\sum_x\hat P_x=\hat I$ and $\hat P_x\hat P_{x'}=\delta_{xx'}\hat P_x$. For a state $\rho$, the probability of outcome $x$ is $p_x=\tr[\hat P_x\rho]$. We call $\rho$ a macroscopic state with respect to $\mathcal C$ if
\begin{align}
    \rho=\sum_x p_x\frac{\hat P_x}{V_x},
    \qquad
    V_x=\tr[\hat P_x].
\end{align}
For an arbitrary state $\rho$, the associated coarse-grained state is
\begin{align}
    \rho_{\mathcal C}:=\sum_x \tr[\hat P_x\rho]\frac{\hat P_x}{V_x}.
\end{align}
One Kraus representation of this map is obtained by choosing an orthonormal basis $\{\ket{x,a}\}_{a=1}^{V_x}$ of the image of $\hat P_x$ and setting
\begin{align}
    K_{x,a,b}=\frac{1}{\sqrt{V_x}}\ketbra{x,a}{x,b},
    \qquad a,b=1,\ldots,V_x.
\end{align}
This representation is used only to make explicit that $\rho_{\mathcal C}$ can be realized as an unselective measurement outcome.
}

\begin{figure}[b]
   \centering
    \includegraphics[width=1\linewidth]{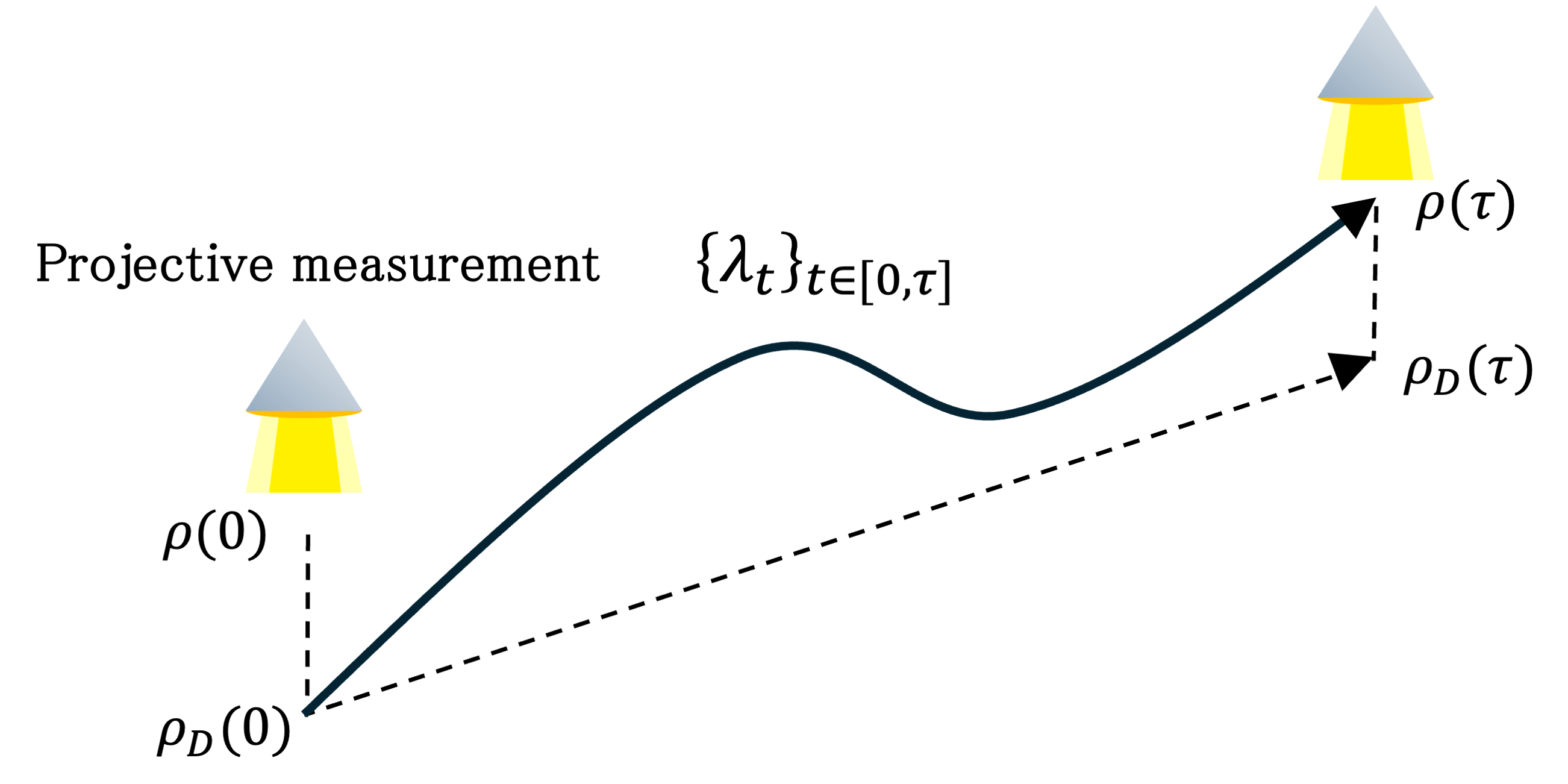}
    \caption{ Schematic of the two-point coarse-graining protocol for an isolated quantum system.}
    \label{FIG-Sche}
\end{figure}

{
We now specify the protocol used throughout the paper. The system is initially prepared with control parameter $\lambda_0$ and state $\rho(0)$. Immediately before the unitary evolution, we apply the coarse-graining measurement associated with the initial PVM $\mathcal C_i=\{\hat P_x\}_x$, so that the initial state becomes $\rho_{\mathcal C_i}(0)$. The system then evolves unitarily during $0\leq t\leq\tau$ under the externally controlled Hamiltonian $H(\lambda_t)$. At the end of the protocol, the final PVM $\mathcal C_f=\{\hat Q_y\}_y$ is applied, yielding
\begin{align}
    \rho_{\mathcal C_f}(\tau)
    =\sum_y q_y\frac{\hat Q_y}{\tr[\hat Q_y]},
    \qquad
    q_y=\tr[\hat Q_y\rho(\tau)].
\end{align}
This two-point coarse-graining framework allows us to compare entropy-related quantities at the beginning and at the end of the unitary protocol; see Fig.~\ref{FIG-Sche}.
}

\section{Thermodynamic entropies and quantum transport distance}
\label{sec:ent-wass}

\subsection{Entropy production in isolated quantum systems}

{
Understanding thermodynamic irreversibility from the reversible dynamics of isolated systems is a central problem in statistical mechanics. The microscopic time evolution of an isolated quantum system is unitary and therefore preserves the fine-grained von Neumann entropy,
\begin{align}
    S_{\rm vN}(\rho(t))=-\tr[\rho(t)\ln\rho(t)]
    =S_{\rm vN}(\rho(0)).
\end{align}
This conservation seems, at first sight, to be in tension with the thermodynamic description of many-body systems, where relaxation, thermalization, and entropy increase are observed as irreversible phenomena.

One resolution is that thermodynamic irreversibility is not a property of the full microscopic state alone, but of a coarse-grained description of that state. Thermodynamics does not track all microscopic degrees of freedom. Instead, it describes a many-body system in terms of macroscopic variables such as energy, particle number, external control parameters, and local conserved quantities. Many distinct microscopic states are therefore identified as the same thermodynamic macrostate. From this viewpoint, irreversibility appears when microscopic information becomes irrelevant or inaccessible within the chosen coarse-grained description, even though the underlying unitary dynamics remains reversible.

For this reason, entropy production in isolated quantum systems should be defined at the same level of description as the relevant macroscopic variables. In what follows, we introduce two such entropy functionals: the Gibbs entropy and observational entropy. A PVM will be used as the mathematical representation of a coarse-graining because it partitions the Hilbert space into mutually orthogonal macrospaces and identifies microscopic states that yield the same macroscopic observable values as the same coarse-grained macrostate.
}

\subsection{Gibbs Entropy}\label{sec:Pss}

The Gibbs entropy is one of the most fundamental entropies that describes thermodynamic properties in equilibrium and satisfies an entropy-increase law under thermodynamic operations. In equilibrium at energy $E$ and control parameter $\lambda$, the Gibbs entropy is defined by
{
\begin{align} 
    \label{def-Sg-equil}
    S_G(E, \lambda) := k_B \ln \tr \Theta(E - H(\lambda)),
\end{align}
where 
\begin{align} 
    \Theta(E-H(\lambda)) := \sum_{i:\, E_i(\lambda)\le E} \hat{P}_i^{(E)}(\lambda)
\end{align}
}
denotes the spectral projection of $H(\lambda)$ onto the subspace of the Hilbert space spanned by eigenstates with eigenvalues not exceeding $E$. In the following, we set $k_B := 1$. In thermodynamically large systems, the Gibbs entropy is known to asymptotically coincide with the Boltzmann entropy.

To examine the validity of the present entropy, we first focus on the fundamental thermodynamic relations that thermodynamic entropy must satisfy.
In thermodynamics, the entropy is defined as a function of the energy and the volume, and it is required to satisfy
\begin{align}
\label{funda-themo}
\pdv{S}{E} = \frac{1}{T}, \quad \pdv{S}{V} = \frac{p}{T},
\end{align}
where $V$, $p$, and $T$ denote the system volume, the thermodynamic pressure, and the absolute temperature, respectively. We next show that an analogous structure also holds for the Gibbs entropy $S_G$ adopted in this work. Introducing the ``Gibbs temperature" $T_G$ and the microscopic pressure $p^{\text{mic}}_{\lambda}$ through
\begin{align}
\label{def-Tg}
\frac{1}{T_G} := \pdv{S_G}{E},\quad p^{\text{mic}}_{\lambda} := -\tr\left[ \pdv{H(\lambda)}{\lambda} \rho^{\text{mic}}(E,\lambda) \right],
\end{align}
we find that
\begin{align}
\pdv{S_G}{\lambda} = \frac{p^{\text{mic}}_{\lambda}}{T_G}
\end{align}
holds. Here, the microcanonical state $\rho^{\text{mic}}(E,\lambda)$ and the density of states $W(E,\lambda)$ are defined as
\begin{align}
&\rho^{\text{mic}}(E,\lambda) := \sum_{i=1}^d \frac{\delta(E_i(\lambda) - E)}{W(E,\lambda)} \ketbra{i_{\lambda}},\\
&W(E,\lambda) := \sum_{i=1}^d \delta(E_i(\lambda) - E),
\end{align}
respectively.
Thus, $S_G$ satisfies differential relations analogous to those of the conventional thermodynamic entropy, making it a natural definition in this sense. {However, it should be noted that} the above reasoning implicitly assumes the validity of $T_G(E)$ itself.
When examining this validity, one must consider issues such as the possibility of negative temperature and the requirement that two systems in thermal contact at equilibrium share the same temperature.
These matters are closely related to—and often discussed together with—the consistency between the Gibbs entropy and the Boltzmann entropy, which represent distinct microscopic formulations of thermodynamic entropy (see e.g., \cite{Dunkel2014,SwendsenWang2015, Braun2013, Ramsey1956}).

{
We emphasize that the Gibbs entropy possesses an important property relevant for its extension to nonequilibrium states—namely, the thermodynamic second law—as will be discussed below. In this paper, we consider 
\begin{align} 
    \label{def-Sg}
    &S_G(\rho,H(\lambda)) := \sum_{i=1}^n \sum_{k=1}^{d_i} \ln(n_{i-1} + k) \f{\tr[\hat{P}_i^{(E)}(\lambda)\rho]}{\tr[\hat{P}_i^{(E)}(\lambda)]},\\
    &n_i = n_{i-1}+d_i, \quad d_i = \tr[\hat{P}_{i}^{(E)}(\lambda)], \quad \quad n_0 = 0
\end{align}
an extension of this definition that remains well defined for nonequilibrium states as well. By choosing $\rho$ as the microcanonical state $\rho^{\text{mic}}(E,\lambda)$, the definition becomes consistent with that of the equilibrium state. 

When the initial state $\rho$ is passive, namely when $\rho$ is diagonal in the eigenbasis $\{\ket{i(\lambda_0)}\}_{i=1}^{n}$ of the Hamiltonian $H(\lambda_0)$ and satisfies for 
\begin{align} 
    \f{\tr[\hat{P}^{(E)}_i(\lambda)[\rho]]}{\tr[\hat{P}^{(E)}_i(\lambda)]} \le \f{\tr[\hat{P}^{(E)}_j(\lambda)[\rho]]}{\tr[\hat{P}^{(E)}_j(\lambda)]}
\end{align}
$\forall i>j$, the second-law-like relation 
 \begin{align} 
    S_G(U\rho U^{\dagger},H(\lambda_{\tau})) \ge S_G(\rho ,H(\lambda_0)),
\end{align}
holds \cite{tasaki2000}. Note that $E_i(\lambda)$ is the $i$-th energy eigenvalue in increasing order. For any passive state, the unitary operator that achieves the equality is, up to irrelevant phase degrees of freedom, uniquely given by $U_{\text{ad}} := \sum_{n=1}^d \ketbra{n_{\lambda_{\tau}}}{n_{\lambda_0}}$ (see Appendix~\ref{app-GibbsUni}). }

\subsection{Observational entropy and thermodynamic coarse-grainings}

\begin{figure*}
    \centering
    \includegraphics[width=0.99\linewidth]{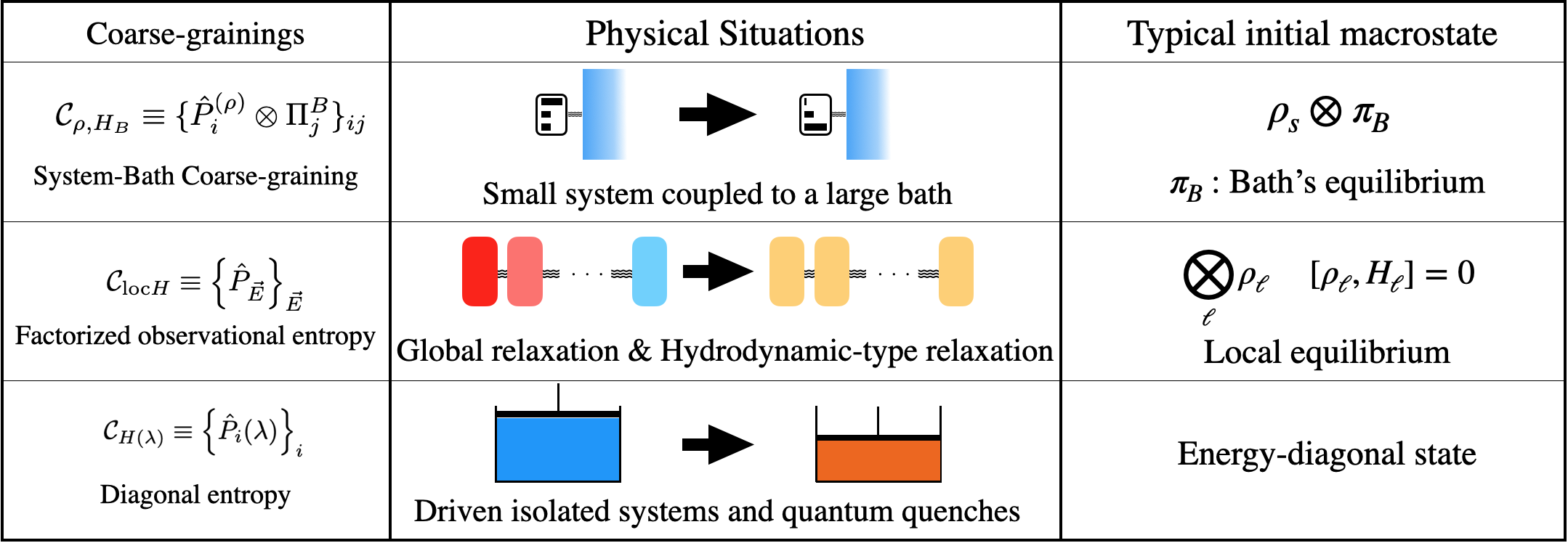}
    \caption{ Thermodynamic coarse-grainings used for observational entropy. The system--bath coarse-graining resolves the bath energy shell and the eigenbasis of the reduced system state; the local-energy coarse-graining resolves the energy profile over spatial cells; and the energy coarse-graining gives the diagonal entropy.}
    \label{fig:Thermo-obs}
\end{figure*}

{
Although the Gibbs entropy is a fundamental thermodynamic entropy for isolated systems, it is tied to the energy ordering and therefore cannot capture all forms of macroscopic relaxation. For example, relaxation under a fixed Hamiltonian can be visible at the level of local observables even when the Gibbs entropy does not change. To describe entropy production associated with more general thermodynamic macrostates, we use observational entropy.

\paragraph{General definition.}
Let $\mathcal C=\{\hat P_x\}_x$ be a coarse-graining of the Hilbert space. For a state $\rho$, define
\begin{align}
    p_x=\tr[\hat P_x\rho],
    \qquad
    V_x=\tr[\hat P_x].
\end{align}
The observational entropy associated with $\mathcal C$ is
\begin{align}
    S_{\mathrm{obs}}^{\mathcal C}(\rho)
    :=
    -\sum_x p_x\ln\frac{p_x}{V_x}.
\end{align}
Equivalently,
\begin{align}
    S_{\mathrm{obs}}^{\mathcal C}(\rho)
    =-
    \sum_x p_x\ln p_x
    +
    \sum_x p_x\ln V_x.
\end{align}
The first term is the Shannon entropy of the coarse-grained outcome distribution, while the second term is the average logarithmic volume of the corresponding macrospaces. Since coarse-graining discards microscopic information,
\begin{align}
    S_{\mathrm{obs}}^{\mathcal C}(\rho)\geq S_{\mathrm{vN}}(\rho).
\end{align}
If the initial state is already macroscopic with respect to $\mathcal C$, this inequality and the conservation of $S_{\rm vN}$ imply
\begin{align}
\label{obsent-inc}
    S_{\mathrm{obs}}^{\mathcal C}(\rho_t)
    \geq
    S_{\mathrm{obs}}^{\mathcal C}(\rho_0).
\end{align}
Thus the increase of observational entropy reflects the loss of information induced by the macroscopic description, not microscopic irreversibility.

Observational entropy depends on the chosen coarse-graining. Hence, only coarse-grainings associated with thermodynamic observables should be interpreted as thermodynamic entropies. In this work we use the three coarse-grainings summarized in Fig.~\ref{fig:Thermo-obs}.
}

{
\paragraph{Thermodynamic coarse-grainings}
We consider three thermodynamically motivated coarse-grainings. 

Thermodynamic speed limits derived for Lindblad dynamics should admit a unitary system--bath description such as in \cite{shiraishi2021}. We therefore first consider an isolated system decomposed into a small system and a bath. For a small system $S$ coupled to a bath $B$, let
\begin{align}
    \Pi_j^B
    :=
    \mathbf 1_j(H_B),
    \qquad
    \rho_S
    :=
    \tr_B[\rho]
    =
    \sum_i\nu_i\hat P_i^{(\rho_S)},
\end{align}
where $\Pi_j^B$ projects onto the $j$-th masoscopic bath-energy window. We define the system--bath coarse-graining by
\begin{align}
    \mathcal C_{\rho,H_B}
    :=
    \left\{
        \hat P_i^{(\rho_S)}\otimes\Pi_j^B
    \right\}_{ij}.
\end{align}
This coarse-graining resolves the reduced-system eigenspaces and the bath-energy windows. Its relation to the conventional open-system entropy production is discussed in Appendix~\ref{SigmaSobs}.

For an isolated many-body system with
\begin{align}
    H
    =
    \sum_{\ell=1}^{m}H_\ell+H_{\mathrm{int}},
    \qquad
    H_\ell
    =
    \sum_{E_\ell}
    E_\ell\hat P_{E_\ell}^{(\ell)},
\end{align}
we define the local-energy coarse-graining by
\begin{align}
    \mathcal C_{\mathrm{loc}H}
    :=
    \left\{
        \bigotimes_{\ell=1}^{m}
        \hat P_{E_\ell}^{(\ell)}
    \right\}_{\vec E},
    \qquad
    S_{\mathrm{FOE}}(\rho)
    :=
    S_{\mathrm{obs}}^{\mathcal C_{\mathrm{loc}H}}(\rho).
\end{align}
This coarse-graining resolves the spatial energy profile and is appropriate for local and hydrodynamic relaxation.

Finally, for the spectral projectors $\hat P_i^{(E)}(\lambda)$ of $H(\lambda)$, we define
\begin{align}
    \mathcal C_{H(\lambda)}
    :=
    \left\{
        \hat P_i^{(E)}(\lambda)
    \right\}_i,
    \qquad
    S_D(\rho;H(\lambda))
    :=
    S_{\mathrm{obs}}^{\mathcal C_{H(\lambda)}}(\rho).
\end{align}
This coarse-graining resolves the global energy occupations and is appropriate for driven isolated systems and quantum quenches. For a nondegenerate Hamiltonian, $S_D$ reduces to the conventional diagonal entropy. \\

The corresponding physical settings and typical initial macrostates are summarized in Fig~\ref{fig:Thermo-obs}. Detailed definitions and physical interpretations of these three thermodynamic coarse-grainings are provided in Appendix~\ref{app:thermodynamic-coarse-grainings}.
}

\subsection{Quantum Wasserstein distance}\label{sec:wasserstein}
In this paper, we employ the quantum Wasserstein distance (VS-QW) introduced by \cite{VuSaito2023}, defined as
\begin{align}
    \label{def-Ws}
    \mathcal{W}_q(\rho,\sigma) := \frac{1}{2}\min_{V^{\dagger}V=\mathbb{I}}\|V\rho V^{\dagger} - \sigma \|_1.
\end{align}
Here, $\|\rho-\sigma\|_1 := \tr \sqrt{(\rho-\sigma)^{\dagger}(\rho-\sigma)}$. The minimum is taken over all unitary transformations $V$. VS-QW was introduced to extend an important property, which holds for classical fluctuating systems, to open quantum systems described by the Lindblad equation \cite{Lindblad1976, Gorini1976}. In our results, two features of VS-QW play a central role: it admits a closed-form expression in terms of the eigenvalues of a given quantum state, and it is a pseudo-metric [i.e. $\mathcal{W}_q(\rho,\sigma)=0$ is possible for $\rho \neq \sigma$].
{ Since $\mathcal W_q$ depends only on the spectra of the two states, $\mathcal W_q(\rho,\sigma)=0$ means unitary equivalence rather than equality of density matrices.}
Specifically, the closed-form expression reads
\begin{equation}
\label{AppA}
    \mathcal{W}_q(\rho,\sigma) 
    = \frac{1}{2} \sum_{i=1}^d \left| \mu_i(\rho) - \mu_i(\sigma) \right|,
\end{equation}
where $\{\mu_i(\rho)\}_{i=1}^d$ and $\{\mu_i(\sigma)\}_{i=1}^d$ are increasing eigenvalues of $\rho$ and $\sigma$, respectively.

\section{Main Results}
\label{sec:main_results}

{
We now present the generalized second-law inequalities and the resulting thermodynamic speed limits. The main message is that, for suitable coarse-grained entropies, entropy production lower-bounds the Vu--Saito quantum Wasserstein distance between the initial and final coarse-grained states. Dividing this entropy-production bound by the average entropy-production rate gives a lower bound on the operation time.
}

\subsection{Gibbs-entropy bound}
\label{subsec:physical_implications_of_the_bounds}

{
\paragraph{Theorem 1: Gibbs entropy.}
Assume that the initial state $\rho(0)$ is passive with respect to the initial Hamiltonian $H(\lambda_0)$. Define the Gibbs entropy production by
\begin{align}
    \Delta S_G(\tau)
    :=
    S_G(\rho(\tau),H(\lambda_\tau))
    -
    S_G(\rho(0),H(\lambda_0)),
\end{align}
 and define
\begin{align}
    \mathcal W_q^G
    :=
    \mathcal W_q\!
    \left(
        \rho_{\mathcal C_{H(\lambda_\tau)}}(\tau),
        \rho(0)
    \right).
\end{align}
Then
\begin{align}
\label{Gibbs_res}
    \Delta S_G(\tau)
    \geq
    \ln2\,(\mathcal W_q^G)^2.
\end{align}
Consequently, with
\begin{align}
    \langle\sigma_G\rangle_\tau
    :=
    \frac{\Delta S_G(\tau)}{\tau},
\end{align}
we obtain the speed limit
\begin{align}
\label{Gibbs_res_speed}
    \tau
    \geq
    \ln2\,
    \frac{(\mathcal W_q^G)^2}{\langle\sigma_G\rangle_\tau}
    =:\tau_{G,2}.
\end{align}
This bound states that, at a fixed average Gibbs entropy-production rate, a finite redistribution of the energy occupation distribution requires a finite operation time. It is relevant to finite-time driving, quantum quenches, and shortcuts to adiabaticity in isolated quantum systems.

For finite-dimensional systems, a sharper linear bound is also available:
\begin{align}
\label{Gibbs_res_small}
    \Delta S_G(\tau)
    &\geq
    \ln\!\left(\frac{d}{d-1}\right)\mathcal W_q^G,\\
\label{Gibbs_res_small_speed}
    \tau
    &\geq
    \ln\!\left(\frac{d}{d-1}\right)
    \frac{\mathcal W_q^G}{\langle\sigma_G\rangle_\tau}
    =:\tau_{G,1}.
\end{align}
Here $d$ is the Hilbert-space dimension. In particular, for a two-level system this linear bound is always tighter than Eq.~\eqref{Gibbs_res}. The proofs of Eqs.~\eqref{Gibbs_res} and \eqref{Gibbs_res_small} are given in Sec.~\ref{sec:Deriv}.
}

\subsection{Bounds for observational entropy production}
\label{subsec:general_theorem_for_observational_entropy}

\subsubsection{General bound}

{
\paragraph{Theorem 2: Observational entropy.}
Let $\mathcal C^i=\{\hat P_x\}_x$ and $\mathcal C^f=\{\hat Q_y\}_y$ be the initial and final coarse-grainings. Assume that the initial state is macroscopic with respect to $\mathcal C^i$,
\begin{align}
    \rho(0)=\rho_{\mathcal C^i}(0)
    :=
    \sum_xp_x(0)\frac{\hat P_x}{V_x}.
\end{align}
After the unitary evolution $\rho(\tau)=U(\tau)\rho(0)U^\dagger(\tau)$, the final coarse-grained state is
\begin{align}
    \rho_{\mathcal C^f}(\tau)
    :=
    \sum_yq_y(\tau)\frac{\hat Q_y}{V_y}.
\end{align}
The observational entropy production is
\begin{align}
\label{Gen-ent-pro}
    \Delta S_{\mathcal C^i\to\mathcal C^f}(\tau)
    :=
    S_{\mathrm{obs}}^{\mathcal C^f}(\rho(\tau))
    -
    S_{\mathrm{obs}}^{\mathcal C^i}(\rho(0)).
\end{align}
Then
\begin{align}
\label{general-res}
    \Delta S_{\mathcal C^i\to\mathcal C^f}(\tau)
    \geq
    2\mathcal W_q\!\left(
        \rho_{\mathcal C^f}(\tau),
        \rho_{\mathcal C^i}(0)
    \right)^2.
\end{align}
This inequality gives a quantitative form of the entropy-increase law. A final coarse-grained state that is separated from the initial macrostate by a finite VS-QW distance necessarily requires a finite amount of observational entropy production.

Defining the average observational entropy-production rate by
\begin{align}
    \bar\sigma_{\mathcal C^i\to\mathcal C^f}(\tau)
    :=
    \frac{\Delta S_{\mathcal C^i\to\mathcal C^f}(\tau)}{\tau},
\end{align}
we obtain the corresponding speed limit
\begin{align}
\label{general-res2}
    \tau
    \geq
    \frac{
        2\mathcal W_q\!\left(
            \rho_{\mathcal C^f}(\tau),
            \rho_{\mathcal C^i}(0)
        \right)^2
    }{
        \bar\sigma_{\mathcal C^i\to\mathcal C^f}(\tau)
    }.
\end{align}
Thus, if the average entropy-production rate is fixed, a process that changes the macrostate by a finite distance cannot be completed in an arbitrarily short time.

The inequality also gives a compact reachability interpretation. For the operation in which a state is first coarse-grained with respect to $\mathcal C$ and then transformed unitarily, observational entropy is a monotone. Equation~\eqref{general-res} strengthens this monotonicity quantitatively: whenever the reverse transformation is impossible because the final and initial coarse-grained spectra are separated, the entropy ordering becomes strict. In this sense, Eqs.~\eqref{Gibbs_res} and \eqref{general-res} are generalized second-law inequalities.

The same mathematical result can also be applied to non-thermodynamic coarse-grainings. As an example, Appendix~\ref{Non-thermo} discusses the entanglement entropy of bipartite pure states.}

{
\subsubsection{Thermodynamic specializations}

\label{subsec:thermodynamic_specializations}

\paragraph{$\mathcal{C}_{\rho,H_B}$ coarse-graining}
\label{subsubsec:energy_coarse_graining}

We first apply the general inequality to the system--bath coarse-graining $\mathcal{C}_{\rho,H_B}$. As the initial state, we assume
\begin{align}
    \rho(0)
    =
    \rho_s(0)\otimes \pi_B,
    \qquad
    \pi_B
    =
    \frac{e^{-\beta \tilde{H}_B}}{\tilde{Z}_B}.
\end{align}
We define the corresponding observational entropy production, the conventional
open-system entropy production, and the Vu--Saito quantum Wasserstein distance
by
\begin{align}
    \Delta S_{\mathrm{SB}}(\tau)
    &:=
    S_{\mathrm{obs}}^{\mathcal C_{\rho(\tau),H_B}}(\rho(\tau))
    -
    S_{\mathrm{obs}}^{\mathcal C_{\rho(0),H_B}}(\rho(0)),
    \\
    \Sigma_\tau
    &:=
    D\!\left(
        \rho(\tau)
        \,\middle\|\,
        \rho_S(\tau)\otimes\pi_B
    \right),
    \\
    \mathcal W_q^{\mathrm{open}}
    &:=
    \mathcal W_q\!\left(
        \rho_{\mathcal C_{\rho(\tau),H_B}}(\tau),
        \rho_S(0)\otimes\pi_B
    \right),
\end{align}
where $\rho_S(\tau)=\tr_B[\rho(\tau)]$. Here, $D(\cdot\| \cdot)$ denotes the Kullback--Leibler divergence. Under the assumptions above,
$\Sigma_\tau$ is equivalently expressed as
\begin{align}
    \Sigma_\tau
    =
    \Delta S_{\mathrm{vN}}(\rho_S)
    +
    \beta Q_\tau .
\end{align}
The detailed definitions of the system--bath coarse-graining, the bath heat
$Q_\tau$, and the derivation of this equivalence are given in
Appendix~\ref{SigmaSobs}.

Then we obtain
\begin{align}
    \Sigma_{\tau}
    &\ge
    \Delta S_{\mathcal{C}_{\rho(0),H_B}\to
    \mathcal{C}_{\rho(\tau),H_B}}(\tau)
    \ge
    2(\mathcal{W}_q^{\mathrm{open}})^2 .
\end{align}
Equivalently, by introducing the average entropy production rates
\begin{align}
    &\bar{\sigma}_{\mathcal{C}_{\rho(0),H_B}\to
    \mathcal{C}_{\rho(\tau),H_B}}(\tau)
    :=
    \frac{
    \Delta S_{\mathcal{C}_{\rho(0),H_B}\to
    \mathcal{C}_{\rho(\tau),H_B}}(\tau)
    }{\tau},\\
    &\sigma_{\tau}
    :=
    \frac{\Sigma_{\tau}}{\tau},
\end{align}
we obtain the thermodynamic speed-limit inequality
\begin{align}
    \tau
    &\ge
    \frac{2(\mathcal{W}_q^{\mathrm{open}})^2}{
    \bar{\sigma}_{\mathcal{C}_{\rho(0),H_B}\to
    \mathcal{C}_{\rho(\tau),H_B}}(\tau)
    }\\
    &\ge
    {\frac{2(\mathcal{W}_q^{\mathrm{open}})^2}{\sigma_{\tau}}} .
\end{align}
These inequalities apply to a small system interacting with a large bath. They provide an isolated-system route to open-system thermodynamic speed limits, with the important distinction that the VS-QW is evaluated for the coarse-grained state of the total system, including the bath.\\

\paragraph{Factorized observational entropy}
\label{subsubsec:local_energy_coarse_graining}

Next, we consider the factorized observational entropy associated with the local-energy coarse-graining $\mathcal{C}_{\mathrm{loc}H}$. We assume that the initial state is given by a product of states that are diagonal with respect to the corresponding local Hamiltonians:
\begin{align}
    \rho(0)
    =
    \bigotimes_{\ell}
    \rho_\ell(0),
    \qquad
    [\rho_\ell(0),H_\ell]=0 .
\end{align}
We denote the corresponding factorized observational entropy production and the Vu--Saito quantum Wasserstein distance by
\begin{align}
    &\Delta S_{\mathrm{FOE}}(\tau)
    :=
    S_{\mathrm{FOE}}(\rho(\tau))
    -
    S_{\mathrm{FOE}}(\rho(0)),\\
    &\mathcal W_q^{\mathrm{FOE}}
    :=
    \mathcal W_q
    \left(
        \rho_{\mathcal{C}_{\mathrm{loc}H}}(\tau),
        \rho_{\mathcal{C}_{\mathrm{loc}H}}(0)
    \right).
\end{align}

Substituting $\mathcal{C}^i=\mathcal{C}^f=\mathcal{C}_{\mathrm{loc}H}$ into the general inequality, we obtain
\begin{align}
    \label{res_foe}
    \Delta S_{\mathrm{FOE}}(\tau)
    \ge
    2(\mathcal W_q^{\mathrm{FOE}})^2 .
\end{align}
Equivalently, by introducing the average production rate
\begin{align}
    \bar{\sigma}_{\mathrm{FOE}}(\tau)
    :=
    \frac{\Delta S_{\mathrm{FOE}}(\tau)}{\tau},
\end{align}
we obtain the thermodynamic speed-limit inequality
\begin{align}
    \label{FOE_speed}
    \tau
    \ge
    \frac{2(\mathcal W_q^{\mathrm{FOE}})^2}{\bar{\sigma}_{\mathrm{FOE}}(\tau)} .
\end{align}
This inequality bounds how fast an isolated many-body system can change its local-energy profile at a fixed rate of factorized observational entropy production. It can therefore be regarded as a finite-size precursor of a hydrodynamic relaxation bound.

If the relaxation time of a hydrodynamic mode is defined as the time required for the initial coarse-grained local-energy distribution to reach the state in which the mode becomes spatially uniform, the FOE speed limit directly gives a lower bound on this relaxation time. Let the initial FOE macrostate and the uniform FOE macrostate be denoted by
\begin{align}
\rho_{\mathrm{FOE}}^{\mathrm{ini}},
\qquad
\rho_{\mathrm{FOE}}^{\mathrm{unif}} .
\end{align}
Then the relaxation time satisfies
\begin{align}
\tau_{\mathrm{rel}}
\ge
\frac{
2\left(\mathcal W_q^{\mathrm{FOE}}\right)^2
}{\bar{\sigma}_{\mathrm{FOE}}(\tau_{\textrm{rel}})
},
\end{align}
where
\begin{align}
\mathcal W_q^{\mathrm{FOE}}
:=
\mathcal W_q
\left(
\rho_{\mathrm{FOE}}^{\mathrm{unif}},
\rho_{\mathrm{FOE}}^{\mathrm{ini}}
\right).
\end{align}
This inequality shows that a hydrodynamic mode whose initial profile is separated from the uniform profile by a finite FOE Wasserstein distance cannot relax in an arbitrarily short time, unless the average FOE entropy-production rate becomes correspondingly large. }

\paragraph{Diagonal entropy}
\label{subsubsec:system_bath_coarse_graining}

Finally, we consider the diagonal entropy associated with the energy coarse-graining $\mathcal{C}_{H(\lambda)}$. In this specialization, we assume that the initial state is diagonal in the energy eigenbasis of the initial Hamiltonian. We abbreviate the corresponding diagonal entropy production and the Vu--Saito quantum Wasserstein distance as
\begin{align}
    \Delta S_D(\tau)
    &:=
    S_D(\rho(\tau); H(\lambda_\tau))
    -
    S_D(\rho(0); H(\lambda_0)),\\
    \mathcal W_q^D
    &:=
    \mathcal W_q
    \left(
        \rho_{\mathcal{C}_{H(\lambda_\tau)}}(\tau),
        \rho_{\mathcal{C}_{H(\lambda_0)}}(0)
    \right).
\end{align}

Substituting \(\mathcal{C}^i=\mathcal{C}_{H(\lambda_0)}\) and \(\mathcal{C}^f=\mathcal{C}_{H(\lambda_\tau)}\) into the general inequality, we obtain
\begin{align}
    \label{Dia_res}
    \Delta S_D(\tau)
    \ge
    2(\mathcal W_q^D)^2 .
\end{align}
Equivalently, by introducing the average diagonal entropy production rate
\begin{align}
    \bar{\sigma}_D(\tau)
    :=
    \frac{\Delta S_D(\tau)}{\tau},
\end{align}
we obtain the thermodynamic speed-limit inequality
\begin{align}
    \label{Dia_speed}
    \tau
    \ge
    \frac{
        2(\mathcal W_q^D)^2
    }{
        \bar{\sigma}_D(\tau)
    } =: \tau_D.
\end{align}
{This inequality bounds how fast a driven isolated system can change its occupation distribution in the energy eigenbasis at a fixed rate of diagonal entropy production. It is therefore directly relevant to irreversibility under time-dependent Hamiltonians, including quantum quenches and shortcuts to adiabaticity.}

\section{Applications}
\label{sec:APP}

\begin{figure}[t]
    \centering
    \includegraphics[width=0.99\linewidth]{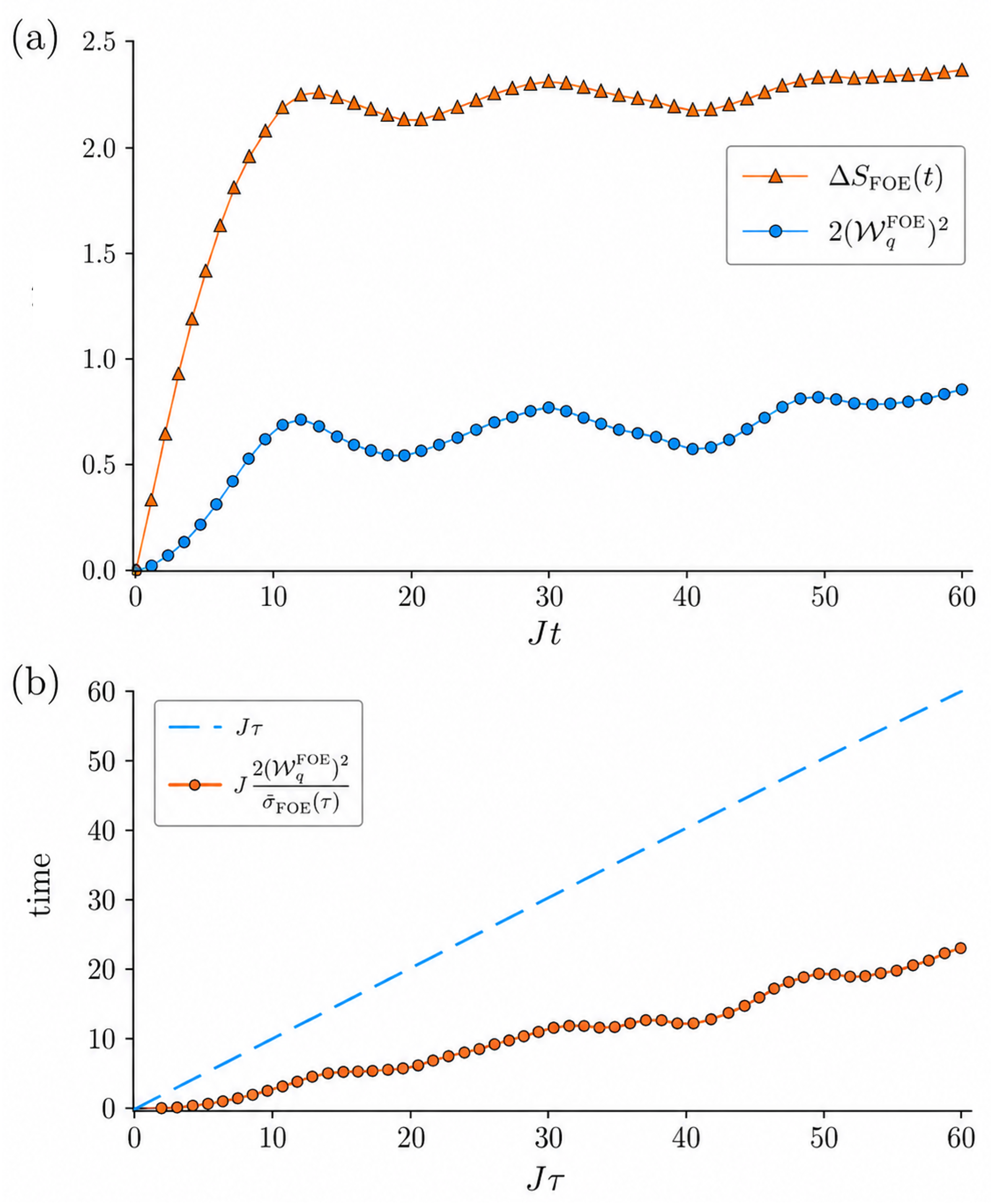}
    \caption{
        {In panel (a), time evolution of the FOE observational entropy production
        $\Delta S_{\mathrm{FOE}}(t)$ and the Wasserstein-distance bound
        $2(\mathcal{W}_q^{\mathrm{FOE}})^2$. In panel (b), thermodynamic speed-limit bound for the FOE relaxation dynamics. The blue dashed line shows the actual elapsed time \(J\tau \), while the orange circles show r.h.s. of Eq.\eqref{FOE_speed}. The orange curve stays below the blue dashed line, confirming the speed-limit inequality.}
        The horizontal axis is $Jt$, and the vertical axis shows the value multiplied by $10^3$. Here, $L = 10, h_x = 1.0, h_z = 1.1, J = 0.01$.
    }
    \label{fig:Wq_FOE}
\end{figure}

{
\subsection{Hydrodynamic-mode relaxation}
To illustrate FOE's bound \eqref{res_foe}, we consider the unitary dynamics of the mixed-field Ising chain
\begin{align}
H
=
-J\sum_{i=1}^{L}\sigma_i^z\sigma_{i+1}^z
-h_x\sum_{i=1}^{L}\sigma_i^x
-h_z\sum_{i=1}^{L}\sigma_i^z ,
\end{align}
with periodic boundary conditions. 
For nonzero $J$, $h_x$, and $h_z$, this model is a simple nonintegrable spin chain.
The initial state is chosen as a product of local Gibbs states,
\begin{align}
\rho(0)
=
\bigotimes_{i=1}^{L}
\frac{e^{-\beta_i H_{\mathrm{loc}}}}{Z_i},
\qquad
H_{\mathrm{loc}}
=
-h_x\sigma^x-h_z\sigma^z ,
\end{align}
{where the inverse temperatures are given by} $\beta_i=1/T_i$, with the site-dependent temperature prepared as a Gaussian profile,
\begin{align}
T_i
=
10
+
10
\exp\left[
-\frac{5(i-\f{L+1}{2})^2}{2L^2}
\right].
\end{align}
The state evolves unitarily as
\begin{align}
\rho(t)=e^{-iHt}\rho(0)e^{iHt}.
\end{align}
We use the local-energy coarse-graining defined by the product eigenbasis of $H_{\mathrm{loc}}$.\\

Figure~\ref{fig:Wq_FOE} shows $\Delta S_{\mathrm{FOE}}(t)$ and $2(\mathcal W_q^{\mathrm{FOE}}(t))^2$. The numerical result confirms that the entropy production is bounded from below by the Wasserstein term throughout the evolution. As shown in Fig.~\ref{fig:Wq_FOE}(a), both quantities start from zero, increase during the initial relaxation, and then approach a slowly varying plateau with finite-size fluctuations. This similarity reflects the fact that both quantities are determined by the same coarse-grained local-energy distribution. The behavior of $2(\mathcal W_q^{\mathrm{FOE}}(t))^2$ therefore suggests that the VS-QW captures information about the coarse-grained local-energy profile in this finite-size example.

Figure~\ref{fig:Wq_FOE}(b) shows the speed-limit lower bound obtained from FOE entropy production. The bound remains below the actual elapsed time $J\tau$, confirming the FOE speed limit throughout the unitary relaxation dynamics. Since the lower bound is finite, the local-energy macrostate cannot change in an arbitrarily short time at a fixed average FOE entropy-production rate. The gap between the bound and the actual time is partly due to the fact that the VS-QW used here is an $O(1)$ quantity, bounded independently of system size. Nevertheless, the bound remains nontrivial for the finite system considered here. The oscillatory behavior reflects finite-size coherent dynamics rather than purely monotonic hydrodynamic diffusion.
}

\subsection{Shortcuts to Adiabaticity}
{The inequalities presented in Sec.~\ref{sec:main_results} show that, in isolated quantum systems, fast state transformations can incur nonadiabatic costs such as Gibbs entropy production and diagonal entropy production.}
A primary example of fast state transformations in isolated systems is quantum computation. In particular, there is an implementation scheme known as adiabatic quantum computation. The computation is carried out by adiabatically evolving the system from the ground state of an initial Hamiltonian, whose ground state can be prepared easily, to the ground state of a final Hamiltonian whose ground state encodes the solution of the problem of interest. Although adiabatic transformations typically require infinitely long operation times, a variety of techniques collectively known as shortcuts to adiabaticity (STA) have been developed to implement them within finite time. Among the various STA methods proposed in the literature, we explain how our main result can be applied to shortcuts to adiabaticity by taking counterdiabatic driving \cite{Demirplak2003, Berry2009} as a concrete example.

Counterdiabatic driving is a method for achieving adiabatic transformations within finite time in isolated quantum systems. For example, consider the case where the ground state $\ket{0(\lambda_0)}$ of the system Hamiltonian $H_0(\lambda_0)$ is driven by external control so that it evolves into the ground state $\ket{0(\lambda_{\tau})}$ of $H_0(\lambda_{\tau})$. However, in general, this type of operation can only be carried out in the quasi-static driving, and thus requires an infinitely long operation time. In counterdiabatic driving, one introduces an additional Hamiltonian that suppresses diabatic transitions
\begin{align}
H_{1}(\lambda_t) := i \dv{\lambda}{t} \sum_{n} \left(\ketbra{\partial_{\lambda}n}{n} - \braket{n}{\partial_{\lambda}n} \ketbra{n}\right),
\end{align}
and modifies the system Hamiltonian to counterdiabatic Hamiltonian $H_0(\lambda_t) + H_{1}(\lambda_t)$. Here, $\ket{n}$ denotes the eigenstates of $H_0$ and $\ket{\partial_{\lambda}n} := \partial_{\lambda}\ket{n}$. With this modification, the quantum state undergoes adiabatic evolution even when $\lambda_t$ is varied rapidly.

From a thermodynamic perspective, it is natural to expect that performing an operation more rapidly entails a thermodynamic cost. {However, STA protocols such as counterdiabatic driving appear to realize finite-time adiabatic operations without endpoint entropy production.} This is because, at the end of the protocol (i.e., when $H_1(\lambda_{\tau}) = 0$), the population distribution over the energy eigenstates is identical to that of the initial state. Consequently, the entropy production vanishes both for the Gibbs entropy and for the diagonal entropy. Therefore, whether a thermodynamic cost associated with the operation time of counterdiabatic driving exists is a nontrivial and important question. Previous studies \cite{Campbell2017,Funoetal2017} have addressed the relation between the operation time and the cost of counterdiabatic driving, but its connection to entropy—the most fundamental quantity characterizing thermodynamic cost—has remained unclear. From Eqs.\eqref{Gibbs_res_speed}, \eqref{Gibbs_res_small_speed} and \eqref{Dia_speed}, we obtain the following inequalities for the operation time $\tau$ of any STA protocol, including counterdiabatic:
\begin{align}
\label{CD-Sg2}
\tau_{\max,2} &:= \ln 2 \max_{0\le t \le \tau}\frac{\mathcal{W}_q(\rho_{\mathcal{C}_{H(\lambda_\tau)}}(\tau),\rho(0))^2}{\langle \sigma_G\rangle_{t}} \le \tau,
\end{align}
\begin{align}
\label{CD-Sg1}
\tau_{\max,1} := \ln \left( \f{d}{d-1} \right) \max_{0\le t \le \tau}\frac{\mathcal{W}_q(\rho_{\mathcal{C}_{H(\lambda_\tau)}}(\tau),\rho(0))}{\langle \sigma_G\rangle_{t}} \le \tau,
\end{align}
\begin{align}
\label{CD-Sd}
\tau_{\max} := 2 \max_{0\le t \le \tau}\frac{\mathcal{W}_q(\rho_{\mathcal{C}_{H(\lambda_\tau)}}(\tau),\rho(0))^2}{\langle \sigma_D \rangle_{t}} \le \tau.
\end{align}
These results follow immediately from the fact that the maxima on the right-hand sides of Eqs.~\eqref{CD-Sg2}–\eqref{CD-Sd} are attained at times smaller than $\tau$.

We explain the meaning of these inequalities. As a remark, the state $\rho_{\mathcal{C}_{H(\lambda_\tau)}}(\tau)$ appearing in Eqs.~\eqref{CD-Sg2}–\eqref{CD-Sd} is the projected state not with respect to the original Hamiltonian $H_0(\lambda_t)$, but with respect to the Hamiltonian under shortcuts to adiabaticity (STA), such as $H_0(\lambda_t) + H_1(\lambda_t)$. {Under the present coarse-grained definition of entropy production, these equations imply that a finite-time STA protocol cannot keep the time-averaged entropy-production rate zero at all intermediate times whenever the intermediate energy-diagonal state is separated from the initial state by a finite VS-QW distance.} This is because, along the protocol, there necessarily exists a time $t\in[0,\tau)$ such that $\mathcal{W}_q(\rho_{\mathcal{C}_{H(\lambda_\tau)}}(\tau),\rho(0))>0$, since $\rho_{\mathcal{C}_{H(\lambda_\tau)}}(\tau)$ corresponds to the quantum state evolving adiabatically at intermediate times. Assuming that the STA protocol is completed at time $\tau$, this implies that $\langle \sigma_D \rangle_{t}>0$ must hold.

We next numerically demonstrate Eqs.~\eqref{CD-Sg2}–\eqref{CD-Sd} using a finite-time Landau–Zener process assisted by a counterdiabatic (CD) protocol \cite{SunZhe2016}.
The counterdiabatic Hamiltonian of this system is given by
\begin{align}
H(t) &= b t\sigma_z + \sigma_x - \frac{b}{2 + 2(b t)^2}\sigma_y,\\
H_0(t) &:= b t\sigma_z + \sigma_x,
\end{align}
where
\begin{align}
\sigma_x =
\begin{pmatrix}
0 & 1\\
1 & 0
\end{pmatrix}, \quad
\sigma_y =
\begin{pmatrix}
0 & i\\
-i & 0
\end{pmatrix}, \quad 
\sigma_z &=
\begin{pmatrix}
1 & 0\\
0 & -1
\end{pmatrix}.
\end{align}
Here, $H_0(t)$ represents the Hamiltonian of the Landau-Zener process, while the term $-b\sigma_y/\{2 + 2(b t)^2\}$ corresponds to the counterdiabatic correction. The operation considered in our numerical calculation consists of the following four steps:
\begin{enumerate}
    \item For time $t\in [-\infty, -b^{-1}]$, the Hamiltonian is fixed as $H_0(-b^{-1}) = -\sigma_z + \sigma_x$.
    \item At $t = -b^{-1}$, we perform a quench to change the Hamiltonian to $H_0(-b^{-1}) - \frac{b}{4}\sigma_y$.
    \item For time $t\in (-b^{-1}, b^{-1})$, the Hamiltonian of the system is given by $H(t)$.
    \item At time $t = b^{-1}$, we perform another quench, changing the Hamiltonian to $H_0(b^{-1}) = \sigma_z + \sigma_x$, after which the operation is terminated.
\end{enumerate}
{In this operation, $b$ determines the operation speed. In the numerical simulations, we varied $b$ and investigated the bounds in Eqs.~\eqref{CD-Sg2}--\eqref{CD-Sd}.} The initial state was chosen as
\begin{align}
\rho = \frac{9}{10}\ketbra{0} + \frac{1}{10}\ketbra{1},
\end{align}
where $\ket{0}$ and $\ket{1}$ denote the ground and excited states of $H_0(-b^{-1})$, respectively.

\begin{figure}[tb]
    \begin{minipage}[b]{0.95\columnwidth}
        \centering
        \includegraphics[width=1\linewidth, angle=0]{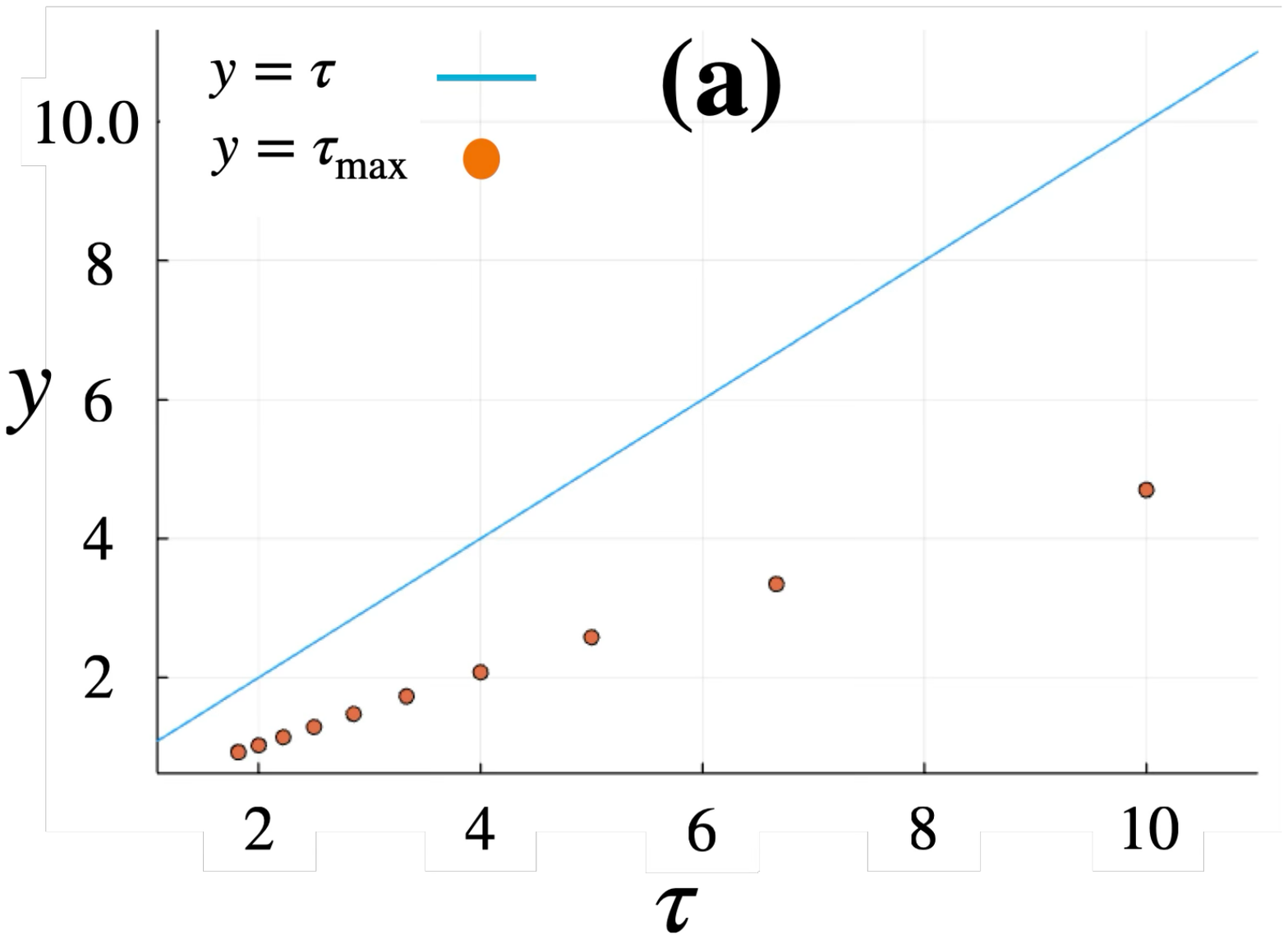}
    \end{minipage}\\
    \begin{minipage}[b]{0.95\columnwidth}
        \centering
        \includegraphics[width=1\linewidth, angle=0]{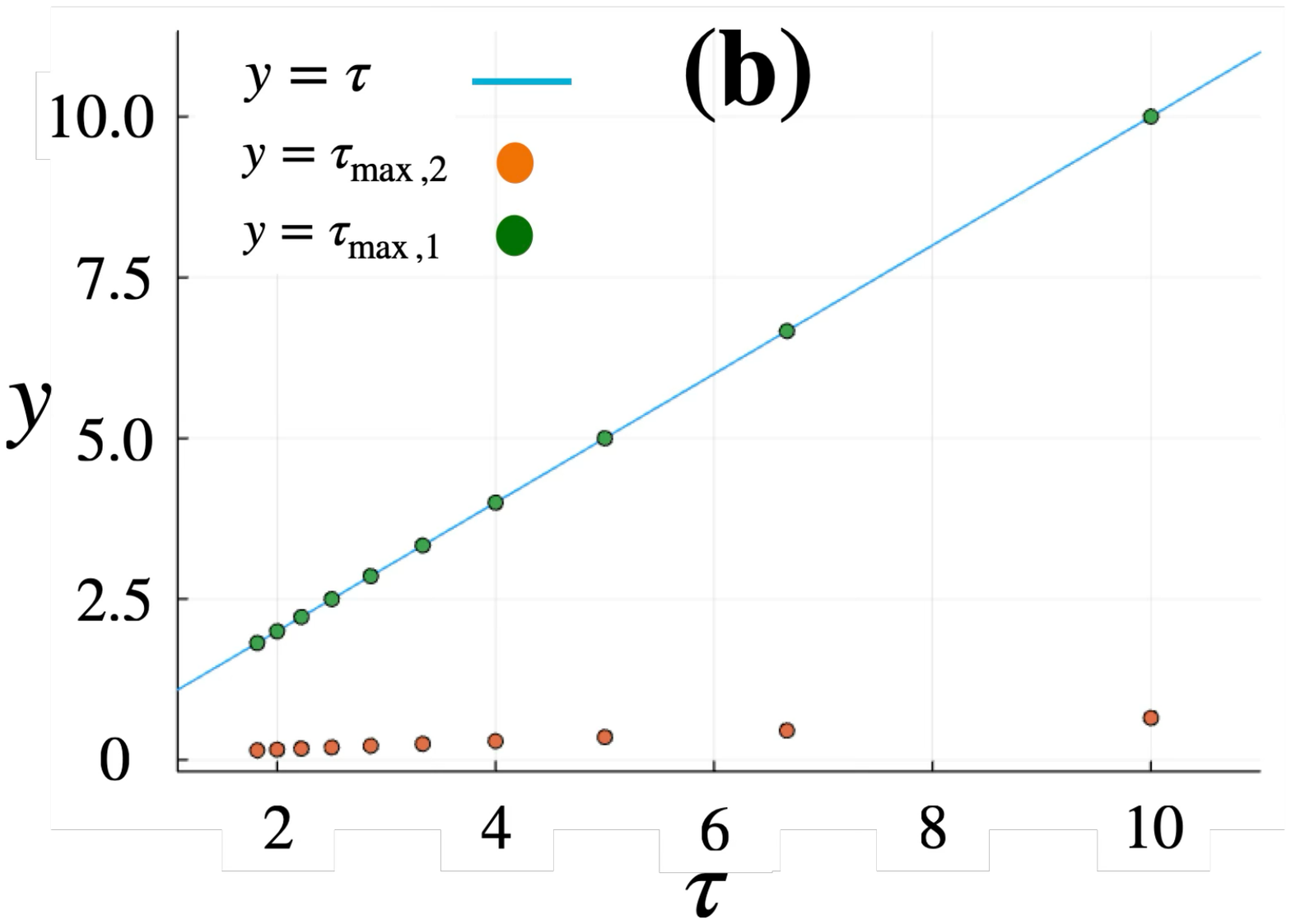}
    \end{minipage}
\protect
\caption{A plot of the LZ counter-diabatic (CD) protocol. We simulated the dynamics of a two-level system driven from the initial Hamiltonian $-10\sigma_z + \sigma_x$ to $10\sigma_z + \sigma_x$ with a counter-diabatic term included, over the time interval $t\in [-t_0, t_0]$ with $t_0 = 10b^{-1}$. The parameter $b$ was varied from $10^3$ to $10^4$ in increments of $10^3$. The horizontal axis shows the operation time, rescaled by a factor of $10^3$. The vertical axis shows the values of the left- and right-hand sides of inequalities Eqs.~\eqref{CD-Sg2}--\eqref{CD-Sd}, also rescaled by $10^3$. In panel (a), the orange dots represent the simulated data corresponding to Eq.~\eqref{CD-Sd}. In panel (b), the orange dots represent the simulated data corresponding to Eq.~\eqref{CD-Sg2}, while the green dots represent the simulated data corresponding to Eq.~\eqref{CD-Sg1}. }
    \label{FIG:t Sg+d}
\end{figure}

{In Fig.~\ref{FIG:t Sg+d}, the dots represent the right-hand sides of Eqs.~\eqref{CD-Sg2}--\eqref{CD-Sd}, while the solid line $y=\tau$ is shown as a reference.
As seen in the figure, for the same operation, the right-hand sides of the inequalities~\eqref{CD-Sg2}--\eqref{CD-Sd} increase monotonically as the operation speed is decreased.

The saturation of the linear Gibbs-entropy bound in Eq.\eqref{Gibbs_res_small_speed} for the present two-level example can be understood as follows. Let \(p_t=(p_g(t),p_e(t))\) be the population distribution of the energy-diagonal state \(\rho_{H(\lambda_t)}(t)\). \(p_e(t)\) denotes the excited-state population. For a two-level system, the Gibbs entropy of the energy-diagonal state is given by \(S_G(\rho_{\mathcal{C}_{H(\lambda_\tau)}}(\tau),H(t))=(\ln 2)p_e(t)\). Since the initial state is passive, the induced doubly stochastic transition gives \(p_e(t)\ge p_e(0)\), and hence the Gibbs entropy production becomes
\begin{align}
\Delta S_G(t)
=
(\ln 2)\bigl[p_e(t)-p_e(0)\bigr]
=
(\ln 2)d_{\rm TV}(p_t,p_{0}),
\end{align}
where \(d_{\rm TV}(p_t,p_{0})=\frac{1}{2}|p_t-p_{0}|_1\). Moreover, as long as \(p_e(t)<1/2\), the ordering of the eigenvalues is unchanged. Therefore, the closed-form expression of the Vu--Saito quantum Wasserstein distance gives
\begin{align}
W_q^D
=
d_{\rm TV}(p_t,p_{0}).
\end{align}
Combining these relations, we obtain
\begin{align}
\Delta S_G(t)
=
(\ln 2)W_q^D,
\end{align}
which shows that the linear bound in Eq.~\eqref{Gibbs_res_small_speed} is saturated.}

\subsection{Rabi cycle: solvable model}

{This example mainly serves as an analytically solvable consistency check of the general bounds. In a shorter version of the paper, it can be moved to an appendix.

We demonstrate our results, Eqs.~\eqref{Gibbs_res}, \eqref{Gibbs_res_speed}, \eqref{Gibbs_res_small}, \eqref{Gibbs_res_small_speed}, \eqref{Dia_res} and \eqref{Dia_speed}. We consider the two-level Hamiltonian
\begin{align}
H(t)
=
\sigma_z
+
b\cos(\omega t)\sigma_x
+
b\sin(\omega t)\sigma_y .
\end{align}
We define
\begin{align}
\Omega
=
\sqrt{
\left(1-\frac{\omega}{2}\right)^2+b^2
},
\qquad
\Omega_0
=
\sqrt{1+b^2}.
\end{align}
The initial state is chosen as the ground state of $H(0)=\sigma_z+b\sigma_x$, namely
\begin{align}
\rho(0)
=
\ket{g(0)}\bra{g(0)} .
\end{align}
Then the excitation probability with respect to the instantaneous energy eigenbasis of $H(t)$ is given by
\begin{align}
P_e(t)
=
\frac{b^2\omega^2}
{4(1+b^2)\left[\left(1-\frac{\omega}{2}\right)^2+b^2\right]}
\sin^2(\Omega t).
\end{align}
VS-QW becomes
\begin{align}
W_q(\rho_{\mathrm{D}}(t),\rho(0))
=
\frac{1-|2P_e(t)-1|}{2}.
\end{align}

The Gibbs entropy production is
\begin{align}
\Delta S_G(t)
=
(\ln 2)P_e(t),
\end{align}
whereas the diagonal entropy production is
\begin{align}
\Delta S_{\mathrm{D}}(t)
=
-
P_e(t)\ln P_e(t)
-
(1-P_e(t))\ln(1-P_e(t)).
\end{align}
We find
\begin{align}
\Delta S_G(t)
=
(\ln 2)W_q(t)
\qquad
\mathrm{for}
\qquad
P_e(t)\leq \frac{1}{2}.
\end{align}
Thus, the linear Gibbs bound is saturated before the eigenvalue-exchange point
\begin{align}
P_e(t)=\frac{1}{2}.
\end{align}
After this point, the Wasserstein distance folds back, while $\Delta S_G(t)$ remains proportional to $P_e(t)$.\\

We briefly discuss the short-time behavior of the quantities introduced above. Since
\begin{align}
\sin^2(\Omega t)
=
\Omega^2t^2+O(t^4),
\end{align}
the excitation probability behaves as
\begin{align}
P_e(t)
=
\gamma t^2+O(t^4),
\qquad
\gamma
:=
\frac{b^2\omega^2}{4(1+b^2)} .
\end{align}
Thus, for sufficiently short times, one has
\begin{align}
P_e(t)<\frac{1}{2}.
\end{align}
In this regime, the ordered eigenvalues of the energy-diagonal state are not exchanged, and the Vu--Saito quantum Wasserstein distance is simply
\begin{align}
W_q(t)
=
P_e(t)
=
\gamma t^2+O(t^4).
\end{align}

The Gibbs entropy production therefore grows quadratically in time:
\begin{align}
\Delta S_G(t)
=
(\ln 2)P_e(t)
=
(\ln 2)\gamma t^2+O(t^4).
\end{align}
In particular, the linear Gibbs bound is saturated in the short-time regime:
\begin{align}
\Delta S_G(t)
=
(\ln 2)W_q(t)
+
O(t^4).
\end{align}
On the other hand, the quadratic Gibbs bound behaves as
\begin{align}
(\ln 2)W_q(t)^2
=
(\ln 2)\gamma^2t^4+O(t^6).
\end{align}
Hence the linear Gibbs bound captures the correct short-time scaling, whereas the quadratic Gibbs bound is subleading.

The diagonal entropy production has a different short-time structure because of the logarithmic singularity of the Shannon entropy near a pure state. Using
\begin{align}
-p\ln p-(1-p)\ln(1-p)
=
p(1-\ln p)+O(p^2),
\end{align}
with
\begin{align}
p=P_e(t),
\end{align}
we obtain
\begin{align}
\Delta S_{\mathrm{D}}(t)
=
\gamma t^2
\left[
1-\ln\left(\gamma t^2\right)
\right]
+
O(t^4|\ln t|).
\end{align}
Therefore, the diagonal entropy production is larger than a purely quadratic function by the logarithmic factor
\begin{align}
|\ln t|.
\end{align}
The corresponding lower bound behaves as
\begin{align}
2W_q(t)^2
=
2\gamma^2t^4+O(t^6).
\end{align}
Thus, for short times, the diagonal-entropy bound is much weaker than the actual diagonal entropy production. This difference originates from the nonanalytic behavior of the Shannon entropy at the initial pure state.}

\begin{figure}[tb]
    \centering
    \includegraphics[width=0.99\linewidth]{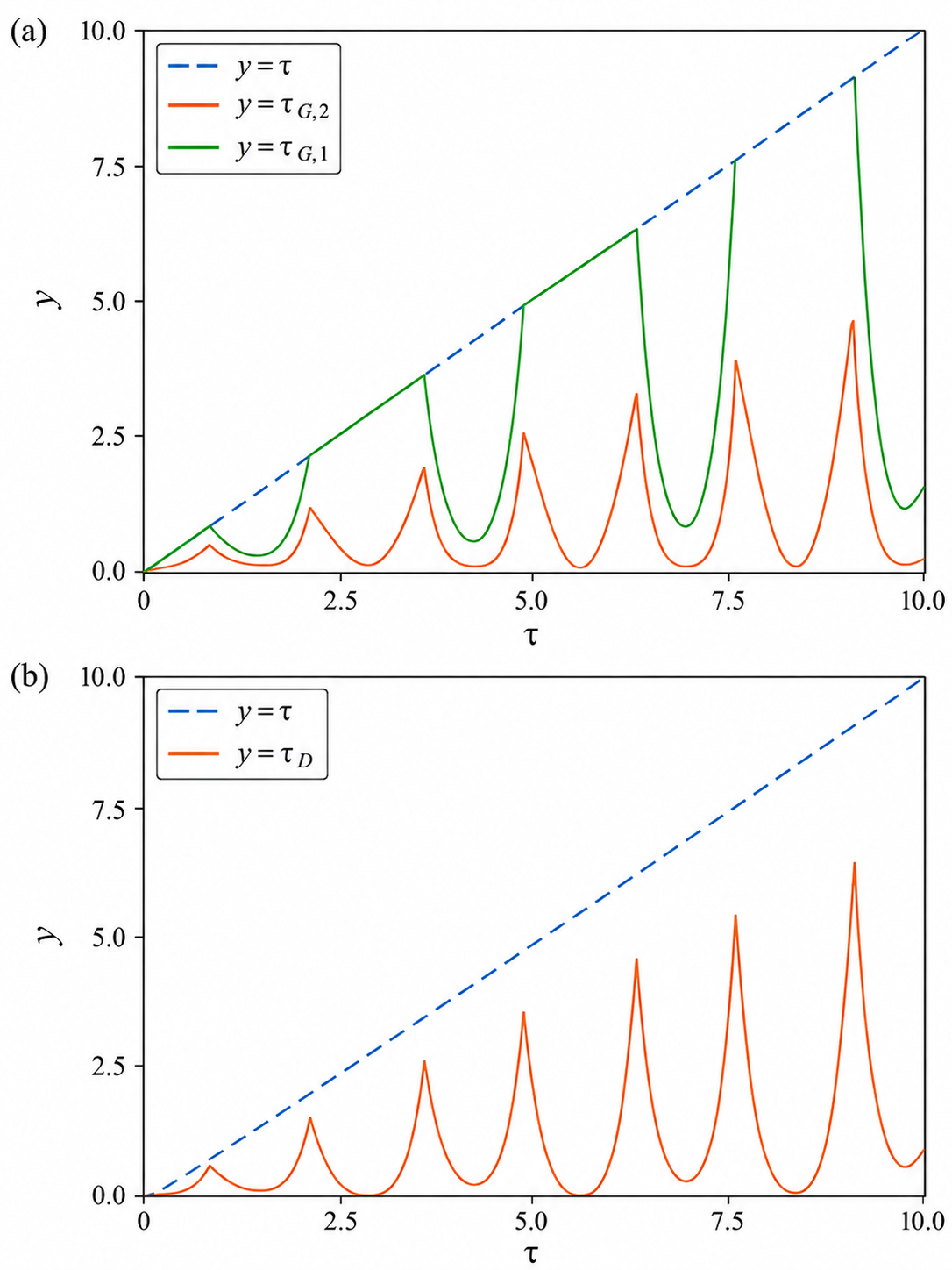}

    \caption{Speed-limit bounds for the Rabi-cycle model. The blue dashed line denotes the actual operation time \(y=\tau\). (a) Gibbs-entropy bounds \(y=\tau_{G,2}\) and \(y=\tau_{G,1}\). (b) Diagonal-entropy bound \(y=\tau_D\). All speed-limit curves lie below \(y=\tau\), confirming the thermodynamic speed-limit inequalities throughout the dynamics.}

    \label{Fig-Rabi-speed}
\end{figure}

\begin{figure}[tb]
    \centering
    \includegraphics[width=0.99\linewidth]{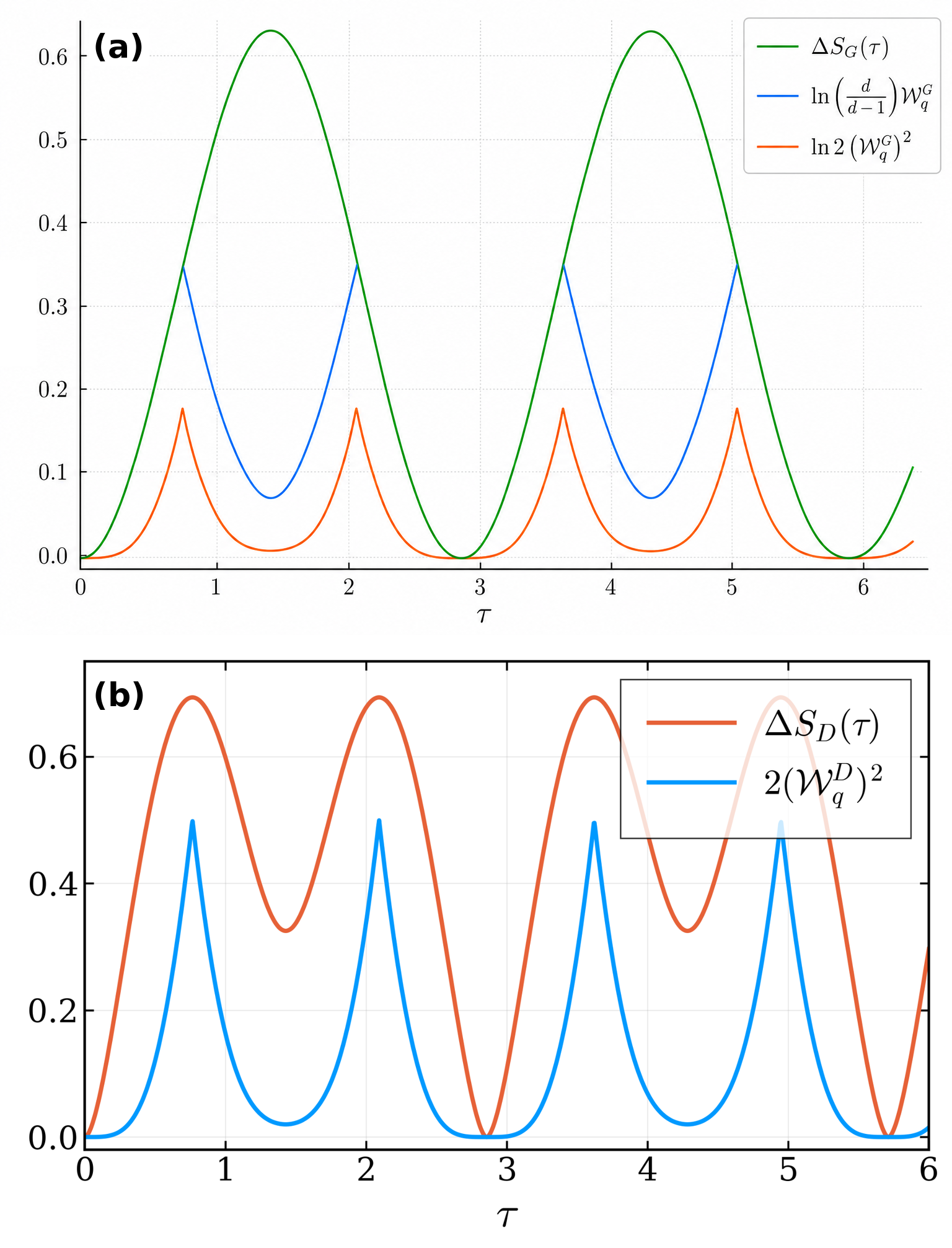}
    \caption{Entropy production and Wasserstein-distance lower bounds in the Rabi-cycle model. (a) Gibbs entropy production $\Delta S_G(\tau)$ together with $\ln\!\left(d/(d-1)\right)\mathcal{W}_q^G$ and $\ln 2\,(\mathcal{W}_q^G)^2$. (b) Diagonal entropy production $\Delta S_D(\tau)$ together with $2(\mathcal{W}_q^D)^2$.}
    \label{Fig-Rabi-second}
\end{figure}

\section{Derivations of Main Results}
\label{sec:Deriv}

{
First, we show the derivation of \eqref{general-res}.
\begin{proof}
We first recall the definitions used below. Let
\(\mathcal C=\{\hat P_x\}_x\) be a PVM on a \(d\)-dimensional Hilbert space,
namely
\begin{align}
    \hat P_x \ge 0, \quad \hat P_x\hat P_y=\delta_{xy}\hat P_x,
    \quad
    \sum_x \hat P_x=\hat 1 .
\end{align}
For a state \(\rho\), the observational entropy associated with
\(\mathcal C\) is defined by
\begin{align}
    S_{\textrm{obs}}^{\mathcal C}(\rho)
    :=
    -\sum_x p_x\ln\frac{p_x}{V_x},
    \quad
    p_x:=\tr[\hat P_x\rho],
    \quad
    V_x:=\tr[\hat P_x].
\end{align}
We say that \(\rho\) is macroscopic with respect to \(\mathcal C\) if
\(S_{\textrm{obs}}^{\mathcal C}(\rho)=S_{\mathrm{vN}}(\rho)\). Equivalently, \(\rho\) is of the
form
\begin{align}
    \rho
    =
    \rho_{\mathcal C}
    :=
    \sum_x p_x\frac{\hat P_x}{V_x}.
\end{align}
In the following, we consider an initial PVM
\(\mathcal C^0=\{\hat P_y^0\}_y\), a final PVM
\(\mathcal C^t=\{\hat P_x^t\}_x\) and a unitary evolution
\(\rho_t=\hat U\rho_0\hat U^\dagger\). We assume that the initial state is
macroscopic with respect to \(\mathcal C^0\), i.e.,
\begin{align}
    \rho_0
    =
    \rho_{\mathcal C^0}
    =
    \sum_y p_y(0)\frac{\hat P_y^0}{V_y^0}.
\end{align}
The final coarse-grained state is
\begin{align}
    \rho_{\mathcal C^t}(t)
    =
    \sum_x p_x(t)\frac{\hat P_x^t}{V_x^t},
    \quad
    p_x(t):=\tr[\hat P_x^t\rho_t],
    \quad
    V_x^t:=\tr[\hat P_x^t].
\end{align}

We now rewrite the observational entropies as Shannon entropies of microscopic
probability vectors. For each macrospace \(x\), introduce \(V_x^t\) microscopic
labels \(i\in x\), and define
\begin{align}
    q_i(t):=\frac{p_x(t)}{V_x^t}
    \qquad (i\in x).
\end{align}
Similarly, for \(j\in y\), define
\begin{align}
    q_j(0):=\frac{p_y(0)}{V_y^0}.
\end{align}
Then
\begin{align}
    &S_{\textrm{obs}}^{\mathcal C^t}(\rho_t)
    =
    -\sum_i q_i(t)\ln q_i(t),\\
    &S_{\textrm{obs}}^{\mathcal C^0}(\rho_0)
    =
    -\sum_j q_j(0)\ln q_j(0).
\end{align}
We next show that \(q(t)\) is obtained from \(q(0)\) by a doubly stochastic
matrix. Indeed, for \(i\in x\),
\begin{align}
    q_i(t)
    &=
    \frac{1}{V_x^t}
    \tr\left[
        \hat P_x^t\hat U\rho_0\hat U^\dagger
    \right]  \nonumber\\
    &=
    \sum_y
    \frac{
    \tr[
        \hat U^\dagger\hat P_x^t\hat U\hat P_y^0
    ]
    }{
    V_x^t V_y^0
    }
    p_y(0) \nonumber\\
    &=
    \sum_y\sum_{j\in y}
    \frac{
    \tr[
        \hat U^\dagger\hat P_x^t\hat U\hat P_y^0
    ]
    }{
    V_x^t V_y^0
    }
    q_j(0).
\end{align}
Thus, if \(i\in x\) and \(j\in y\), we define
\begin{align}
    B_{ij} :=
    \frac{
    \tr[
        \hat U^\dagger\hat P_x^t\hat U\hat P_y^0
    ]}{V_x^t V_y^0},
\end{align}
then
\begin{align}
    q_i(t)=\sum_j B_{ij}q_j(0).
\end{align}
The matrix \(B\) is nonnegative, because for 
\begin{align}
B_{ij}V_x^t V_y^0 &= \tr[\sqrt{P_y^0}\hat U^\dagger\hat P_x^t\hat U\sqrt{\hat P_y^0}] \nonumber\\
&= \tr[\left(\sqrt{P_y^0}\hat U^\dagger\sqrt{\hat P_x^t}\right)\left(\sqrt{P_y^0}\hat U\sqrt{\hat P_x^t}\right)^{\dagger}] \ge 0.
\end{align}
Moreover,
\begin{align}
    \sum_i B_{ij}
    &=
    \sum_x\sum_{i\in x}
    \frac{
    \tr[
        \hat U^\dagger\hat P_x^t\hat U\hat P_y^0
    ]
    }{
    V_x^t V_y^0
    } \nonumber\\
    &=
    \sum_x
    \frac{
    \tr[
        \hat U^\dagger\hat P_x^t\hat U\hat P_y^0
    ]
    }{
    V_y^0
    }
    =
    1,
\end{align}
and
\begin{align}
    \sum_j B_{ij}
    &=
    \sum_y\sum_{j\in y}
    \frac{
    \tr[
        \hat U^\dagger\hat P_x^t\hat U\hat P_y^0
    ]
    }{
    V_x^t V_y^0
    } \nonumber\\
    &=
    \sum_y
    \frac{
    \tr[
        \hat U^\dagger\hat P_x^t\hat U\hat P_y^0
    ]
    }{
    V_x^t
    }
    =
    1.
\end{align}
Therefore \(B\) is doubly stochastic.

Let \(Q(0)\) and \(Q(t)\) be the probability vectors obtained by rearranging
\(q(0)\) and \(q(t)\) in nonincreasing order. Since permutations are also doubly
stochastic, the transition from \(Q(0)\) to \(Q(t)\) is again described by a
doubly stochastic matrix. Hence \(Q(0)\) majorizes \(Q(t)\). Using the Lenard--Tasaki inequality in Appendix~\ref{app-q-p}, we obtain 
\begin{align}
    &S_{\textrm{obs}}^{\mathcal C^t}(\rho_t)-S_{\textrm{obs}}^{\mathcal C^0}(\rho_0) \nonumber\\
    &=
    -\sum_i Q_i(t)\ln Q_i(t)
    +
    \sum_i Q_i(0)\ln Q_i(0) \nonumber\\
    &=
    D\!\left(Q(0)\middle\|Q(t)\right)
    +
    \sum_i
    \bigl(Q_i(0)-Q_i(t)\bigr)\ln Q_i(t) \nonumber\\
    &\ge
    D\!\left(Q(0)\middle\|Q(t)\right).
\end{align}
Here
\begin{align}
    D\!\left(Q(0)\middle\|Q(t)\right)
    :=
    \sum_i Q_i(0)\ln\frac{Q_i(0)}{Q_i(t)}
\end{align}
is the classical relative entropy. Applying Pinsker's inequality gives
\begin{align}
    S_{\textrm{obs}}^{\mathcal C^t}(\rho_t)-S_{\textrm{obs}}^{\mathcal C^0}(\rho_0)
    \ge
    \frac12
    \left\|Q(t)-Q(0)\right\|_1^2 .
\end{align}

It remains to identify this classical distance with VS-QW. The eigenvalue lists of
\(\rho_{\mathcal C^t}(t)\) and \(\rho_{\mathcal C^0}(0)\) are precisely
\(Q(t)\) and \(Q(0)\), respectively. Therefore, by the closed form of VS-QW (see, App.\ref{app-Wq}),
\begin{align}
    \mathcal W_q
    \left(
        \rho_{\mathcal C^t}(t),
        \rho_{\mathcal C^0}(0)
    \right)
    =
    \frac12
    \left\|Q(t)-Q(0)\right\|_1 .
\end{align}
Consequently,
\begin{align}
    S_{\textrm{obs}}^{\mathcal C^t}(\rho_t)-S_{\textrm{obs}}^{\mathcal C^0}(\rho_0)
    \ge
    2
    \mathcal W_q
    \left(
        \rho_{\mathcal C^t}(t),
        \rho_{\mathcal C^0}(0)
    \right)^2 .
\end{align}
This proves the desired entropy-production bound.
\end{proof}
}

Eqs.\eqref{Gibbs_res} and \eqref{Gibbs_res_small}.
\begin{proof}
    Let the spectral decompositions of the initial and final Hamiltonians be
    \begin{align}
        \begin{split}
        \hat H(\lambda_0) =
        \sum_{i=1}^{n_0}
        E_i(\lambda_0) \hat P_i^{(E)}(\lambda_0),\\
        \quad
        \hat H(\lambda_{\tau}) =
        \sum_{i=1}^{n_{\tau}}
        E_i(\lambda_{\tau}) \hat P_i^{(E)}(\lambda_{\tau}).
        \end{split}
    \end{align}
    For the initial and final states $\rho(0)$ and $\rho(\tau)$, consider classical probability distributions $p,q \in \mathbb{R}^d$ whose $i$-th components are
    \begin{align}
        p_i &= \f{\tr[\hat P_i^{(E)}(\lambda_0) \rho(0)]}{\tr[\hat P_i^{(E)}(\lambda_0)]},\quad 
        q_i &= \f{\tr[\hat P_i^{(E)}(\lambda_{\tau}) \rho(\tau)]}{\tr[\hat P_i^{(E)}(\lambda_{\tau})]},
    \end{align}
    As in the case of observational entropy, the matrix \(B\) describing the time evolution from \(\{p_i\}_i\) to \(\{q_i\}_i\), defined by
    \begin{align}
        B_{ij}
        =
        \frac{
            \tr
            \left[
                \hat{U}^{\dagger}
                \hat{P}_i^{(E)}(\lambda_{\tau})
                \hat{U}
                \hat{P}_j^{(E)}(\lambda_0)
            \right]
        }{
            \tr
            \left[
                \hat{P}_j^{(E)}(\lambda_0)
            \right]
            \tr
            \left[
                \hat{P}_i^{(E)}(\lambda_{\tau})
            \right]
        },
    \end{align}
    is a doubly stochastic matrix. Birkhoff-von Neumann theorem \cite{Marshall2011} states that any doubly stochastic matrix can be expressed as a convex combination of permutation matrices. Accordingly, we represent $T$ as $T=\sum_a c_a R_a$, where ${c_a}$ is a discrete probability distribution and $R_a$ are permutation matrices.\\
    Next, let $x\in\mathbb{R}^d$ and consider the following functions
    \begin{align}
        V_1(x) &:= \sum_{n=1}^d \ln n (x_n - p_n) - \f{1}{2}\ln (\frac{d}{d-1}) \|x-p\|_1.\\
        V_2(x) &:= \sum_{n=1}^d \ln n (x_n - p_n) - \frac{\ln2}{4}\|x-p\|_1^2.
    \end{align}
    Here, $\|x-p\|_1=\sum_n |x_n-p_n|$. Since the first term of $V_1(x), V_2(x)$ is concave and the second terms are convex, it follows that $V_1(x), V_2(x)$ are concave. A function $f:\mathbb{R}^d \to \mathbb{R}$ is said to be concave if it satisfies the following condition:$\forall x,y \in \mathbb{R}^d, \forall \lambda \in [0,1],$
	\begin{align}
		f(\lambda x+(1-\lambda)y) \ge \lambda f(x) + (1-\lambda)f(y).
	\end{align}
    Therefore, we obtain 
    \begin{align}
        \label{V>V_a}
        V_i(Tp) \ge \sum_a c_a V_i(R_ap), \quad (i=1,2).
    \end{align}
    Focus on one permutation matrix $R_a$ contained in $B$. If $R_a$ permutes the $n$-th and $m$-th components with $n>m$, then we obtain
    \begin{align}
        \label{V_a>m-n}
        V_1(R_ap) &= (p_m-p_n)\left\{ \ln \frac{n}{m} - \ln (\f{d}{d-1})\right\},\\
        \label{V_a>m-n2}
        V_2(R_ap) &= (p_m-p_n)\left( \ln \frac{n}{m} - \ln2(p_m - p_n)\right).
    \end{align}
    Since $p_m \ge p_n$, the signs of $V_1(R_ap)$ and $V_2(R_ap)$ are determined by $\ln (n/m) - \ln \{d/(d-1)\}$ and $\ln (n/m) - (\ln2) (p_m - p_n)$, respectively. Because $\ln (n/m) \ge \ln \{(m+1)/m\} \ge \ln \{d/(d-1)\}$, we obtain $V_1(R_ap) \ge 0$. From the passivity of $\rho(0)$ and normalization constant, one has $1/m \ge p_m$; together with $n>m$, this implies
    \begin{align}
        \label{V_a>0}
        \ln \frac{n}{m} - \ln2(p_m - p_n) \ge \ln \frac{m+1}{m} - \ln2 \frac{1}{m}.
    \end{align}
    The right-hand side of Eq.~\eqref{V_a>0} is positive for $m>0$, as can be verified by a straightforward calculation. Combining Eqs.~\eqref{V>V_a}, \eqref{V_a>m-n}, \eqref{V_a>m-n2} and \eqref{V_a>0}, we obtain $V_1(Tp)\ge 0, V_2(Tp)\ge 0$. Therefore, we obtain
    \begin{align}
       &S_G(\rho(\tau),\lambda_{\tau}) - S_G(\rho(0),\lambda_0) \ge \f{1}{2}\ln (\frac{d}{d-1}) \|q-p\|_1.\\
        &S_G(\rho(\tau),\lambda_{\tau}) - S_G(\rho(0),\lambda_0) \ge \frac{\ln2}{4}\|q-p\|_1^2.
    \end{align}

    Next, we relate the classical distance \(\|q-p\|_1\) to the Vu--Saito quantum Wasserstein distance. We consider a unitary \(U\) that maps the \(n\)-th energy eigenstate of \(H(\lambda_{\tau})\) to the \(n\)-th energy eigenstate of \(H(\lambda_{0})\). If degeneracies are present, we use an arbitrarily fixed orthonormal basis in each degenerate eigenspace.Then, we have
    \begin{align}
        \frac{1}{2}\|q-p\|_1
        =
        \frac{1}{2}
        \left\|
            U\rho_{H(\lambda_\tau)}(\tau)U^{\dagger}
            -
            \rho(0)
        \right\|_1 .
    \end{align}
    By the definition of the Vu--Saito quantum Wasserstein distance, we further obtain
    \begin{align}
        \frac{1}{2}\|q-p\|_1
        &=
        \frac{1}{2}
        \left\|
            U\rho_{H(\lambda_\tau)}(\tau)U^{\dagger}
            -
            \rho(0)
        \right\|_1 \nonumber\\
        &\ge
        \mathcal{W}_q
        \left(
            U\rho_{H(\lambda_\tau)}(\tau)U^{\dagger},
            \rho(0)
        \right).
    \end{align}

    We conclude from the above that
    \begin{align}
        S_G(\rho(\tau),\lambda_{\tau}) - S_G(\rho(0),\lambda_0) &\ge \ln \f{d}{d-1} \mathcal{W}_q(\rho_D(\tau),\rho(0)),\\
        S_G(\rho(\tau),\lambda_{\tau}) - S_G(\rho(0),\lambda_0) &\ge \ln 2 \mathcal{W}_q(\rho_D(\tau),\rho(0))^2.
    \end{align}
\end{proof}

\section{Concluding remarks and outlooks}
\label{sec:remarks}

{
In this work, we derived thermodynamic speed limits for isolated quantum systems undergoing finite-time unitary dynamics. Since the von Neumann entropy is conserved under unitary evolution, thermodynamic irreversibility must be formulated at the level of coarse-grained descriptions. We showed that entropy production associated with the Gibbs entropy and with several thermodynamically motivated observational entropies is bounded from below by the Vu--Saito quantum Wasserstein distance between the initial and final coarse-grained states. Dividing these generalized second-law inequalities by the average entropy-production rate gives lower bounds on the operation time.

The resulting bounds clarify a thermodynamic aspect of speed limits in isolated quantum systems. Ordinary quantum speed limits constrain the operation time by kinematic quantities, such as Hilbert-space distances and energetic fluctuations. By contrast, the bounds derived here constrain the operation time by the entropic cost required to change a thermodynamic macrostate. They therefore quantify the statement that a finite change of a coarse-grained state cannot be achieved in an arbitrarily short time under a fixed entropy-production rate. We also note that the present lower bounds are complementary to standard continuity bounds for entropy. Since observational entropy is the von Neumann entropy of the corresponding coarse-grained state, Fannes–Audenaert-type inequalities \cite{CoverThomas2006, NielsenChuang2010, Audenaert2007} give upper bounds
\begin{align}
    \Delta S_{\mathcal C^i\to\mathcal C^f}(\tau) \le T \log_2 (d-1) + H((T,1-T))
\end{align}
on the entropy production in terms of the trace distance $T = \f{1}{2}\|\rho_{\mathcal C^f}(\tau) - \rho_{\mathcal C^i}(0)\|_1$ between the initial and final coarse-grained states. Here, $H((p, 1-p))$ denotes the binary entropy function. Thus, coarse-grained entropy production is constrained from both below and above by distances between macroscopic states.

A central open problem is to construct an extensive version of the quantum Wasserstein distance suitable for macroscopic systems. The Vu--Saito distance used here is an $O(V^0)$ quantity, so the present lower bounds may become weak when entropy production scales extensively with system size. Developing hydrodynamic or extensive quantum transport distances would turn the finite-dimensional inequalities derived here into thermodynamically robust speed limits for macroscopic isolated quantum systems. Another important direction is to clarify how the present isolated-system bounds reduce to, or differ from, known open-system speed limits under weak-coupling, large-bath, or Markovian limits.
}

\begin{acknowledgments}
RK is grateful to Keiji Saito for his continuous discussions and encouragement. We thank Kazuma Yokota and Masato Itami, Kyota Tamano for careful reading of the manuscript and helpful comments. 
This study was supported by JST SPRING Grant Number JPMJSP2110 and Graduate School of Science, Kyoto University under Ginpu Fund,  2025. We are also supported by JSPS KAKENHI Grant No. JP23K25796, No. JP26H02015, and JP26H00388.

\end{acknowledgments}

\twocolumngrid
\bibliography{refs}

@article{Yung2006,
  title = {Quantum speed limit for perfect state transfer in one dimension},
  author = {Yung, Man-Hong},
  journal = {Phys. Rev. A},
  volume = {74},
  issue = {3},
  pages = {030303},
  numpages = {4},
  year = {2006},
  month = {Sep},
  publisher = {American Physical Society},
  doi = {10.1103/PhysRevA.74.030303},
  url = {https://link.aps.org/doi/10.1103/PhysRevA.74.030303}
}

@article{Caneva2009,
  title = {Optimal Control at the Quantum Speed Limit},
  author = {Caneva, T. and Murphy, M. and Calarco, T. and Fazio, R. and Montangero, S. and Giovannetti, V. and Santoro, G. E.},
  journal = {Phys. Rev. Lett.},
  volume = {103},
  issue = {24},
  pages = {240501},
  numpages = {4},
  year = {2009},
  month = {Dec},
  publisher = {American Physical Society},
  doi = {10.1103/PhysRevLett.103.240501},
  url = {https://link.aps.org/doi/10.1103/PhysRevLett.103.240501}
}

@article{JonesKok2010,
  title = {Geometric derivation of the quantum speed limit},
  author = {Jones, Philip J. and Kok, Pieter},
  journal = {Phys. Rev. A},
  volume = {82},
  issue = {2},
  pages = {022107},
  numpages = {7},
  year = {2010},
  month = {Aug},
  publisher = {American Physical Society},
  doi = {10.1103/PhysRevA.82.022107},
  url = {https://link.aps.org/doi/10.1103/PhysRevA.82.022107}
}

@Inbook{Mandelstam1991,
author="Mandelstam, L.
and Tamm, Ig.",
editor="Bolotovskii, Boris M.
and Frenkel, Victor Ya.
and Peierls, Rudolf",
title="The Uncertainty Relation Between Energy and Time in Non-relativistic Quantum Mechanics",
bookTitle="Selected Papers",
year="1991",
publisher="Springer Berlin Heidelberg",
address="Berlin, Heidelberg",
pages="115--123"
}

@article{MargolusLevitin1998,
title = {The maximum speed of dynamical evolution},
journal = {Physica D: Nonlinear Phenomena},
volume = {120},
number = {1},
pages = {188-195},
year = {1998},
note = {Proceedings of the Fourth Workshop on Physics and Consumption},
issn = {0167-2789},
doi = {https://doi.org/10.1016/S0167-2789(98)00054-2},
url = {https://www.sciencedirect.com/science/article/pii/S0167278998000542},
author = {Norman Margolus and Lev B. Levitin},
}

@article{Seth2000,
  author  = {Lloyd, Seth},
  title   = {Ultimate physical limits to computation},
  journal = {Nature},
  year    = {2000},
  month   = aug,
  day     = {31},
  volume  = {406},
  number  = {6799},
  pages   = {1047--1054},
  doi     = {10.1038/35023282},
  url     = {https://doi.org/10.1038/35023282}
}

@article{DeffnerLuts2010,
  title = {Generalized Clausius Inequality for Nonequilibrium Quantum Processes},
  author = {Deffner, Sebastian and Lutz, Eric},
  journal = {Phys. Rev. Lett.},
  volume = {105},
  issue = {17},
  pages = {170402},
  numpages = {4},
  year = {2010},
  month = {Oct},
  publisher = {American Physical Society},
  doi = {10.1103/PhysRevLett.105.170402},
  url = {https://link.aps.org/doi/10.1103/PhysRevLett.105.170402}
}

@article{Deffner2017,
doi = {10.1088/1751-8121/aa86c6},
url = {https://doi.org/10.1088/1751-8121/aa86c6},
year = {2017},
month = {oct},
publisher = {IOP Publishing},
volume = {50},
number = {45},
pages = {453001},
author = {Deffner, Sebastian and Campbell, Steve},
title = {Quantum speed limits: from Heisenberg’s uncertainty principle to optimal quantum control},
journal = {J. Phys. A: Math. Theor.},
}

@article{Aurelletal2011,
  title = {Optimal Protocols and Optimal Transport in Stochastic Thermodynamics},
  author = {Aurell, Erik and Mej\'{\i}a-Monasterio, Carlos and Muratore-Ginanneschi, Paolo},
  journal = {Phys. Rev. Lett.},
  volume = {106},
  issue = {25},
  pages = {250601},
  numpages = {4},
  year = {2011},
  month = {Jun},
  publisher = {American Physical Society},
  doi = {10.1103/PhysRevLett.106.250601},
  url = {https://link.aps.org/doi/10.1103/PhysRevLett.106.250601}
}

@book{Villani2009,
  author    = {Villani, C{\'e}dric},
  title     = {Optimal Transport: Old and New},
  series    = {Grundlehren der mathematischen Wissenschaften},
  volume    = {338},
  publisher = {Springer},
  address   = {Berlin, Heidelberg},
  year      = {2008},
  doi       = {10.1007/978-3-540-71050-9},
  isbn      = {978-3-540-71049-3}
}

@article{BenamouBrenier2000,
  author  = {Benamou, Jean-David and Brenier, Yann},
  title   = {A computational fluid mechanics solution to the Monge-Kantorovich mass transfer problem},
  journal = {Numerische Mathematik},
  year    = {2000},
  volume  = {84},
  number  = {3},
  pages   = {375--393},
  month   = jan,
  doi     = {10.1007/s002110050002},
  url     = {https://link.springer.com/article/10.1007/s002110050002}
}

@article{VuSaito2023,
  title = {Thermodynamic Unification of Optimal Transport: Thermodynamic Uncertainty Relation, Minimum Dissipation, and Thermodynamic Speed Limits},
  author = {Van Vu, Tan and Saito, Keiji},
  journal = {Phys. Rev. X},
  volume = {13},
  issue = {1},
  pages = {011013},
  numpages = {45},
  year = {2023},
  month = {Feb},
  publisher = {American Physical Society},
  doi = {10.1103/PhysRevX.13.011013},
  url = {https://link.aps.org/doi/10.1103/PhysRevX.13.011013}
}

@article{Shiraishifunosaito2018,
  title = {Speed Limit for Classical Stochastic Processes},
  author = {Shiraishi, Naoto and Funo, Ken and Saito, Keiji},
  journal = {Phys. Rev. Lett.},
  volume = {121},
  issue = {7},
  pages = {070601},
  numpages = {6},
  year = {2018},
  month = {Aug},
  publisher = {American Physical Society},
  doi = {10.1103/PhysRevLett.121.070601},
  url = {https://link.aps.org/doi/10.1103/PhysRevLett.121.070601}
}

@article{Ito2018,
  title = {Stochastic Thermodynamic Interpretation of Information Geometry},
  author = {Ito, Sosuke},
  journal = {Phys. Rev. Lett.},
  volume = {121},
  issue = {3},
  pages = {030605},
  numpages = {7},
  year = {2018},
  month = {Jul},
  publisher = {American Physical Society},
  doi = {10.1103/PhysRevLett.121.030605},
  url = {https://link.aps.org/doi/10.1103/PhysRevLett.121.030605}
}

@article{ItoDechant2020,
  title = {Stochastic Time Evolution, Information Geometry, and the Cram\'er-Rao Bound},
  author = {Ito, Sosuke and Dechant, Andreas},
  journal = {Phys. Rev. X},
  volume = {10},
  issue = {2},
  pages = {021056},
  numpages = {27},
  year = {2020},
  month = {Jun},
  publisher = {American Physical Society},
  doi = {10.1103/PhysRevX.10.021056},
  url = {https://link.aps.org/doi/10.1103/PhysRevX.10.021056}
}

@article{Tuanetal2020,
  title = {Unified approach to classical speed limit and thermodynamic uncertainty relation},
  author = {Vo, Van Tuan and Van Vu, Tan and Hasegawa, Yoshihiko},
  journal = {Phys. Rev. E},
  volume = {102},
  issue = {6},
  pages = {062132},
  numpages = {8},
  year = {2020},
  month = {Dec},
  publisher = {American Physical Society},
  doi = {10.1103/PhysRevE.102.062132},
  url = {https://link.aps.org/doi/10.1103/PhysRevE.102.062132}
}

@article{YoshimuraIto2021,
  title = {Thermodynamic Uncertainty Relation and Thermodynamic Speed Limit in Deterministic Chemical Reaction Networks},
  author = {Yoshimura, Kohei and Ito, Sosuke},
  journal = {Phys. Rev. Lett.},
  volume = {127},
  issue = {16},
  pages = {160601},
  numpages = {6},
  year = {2021},
  month = {Oct},
  publisher = {American Physical Society},
  doi = {10.1103/PhysRevLett.127.160601},
  url = {https://link.aps.org/doi/10.1103/PhysRevLett.127.160601}
}

@article{Funo2019,
doi = {10.1088/1367-2630/aaf9f5},
url = {https://doi.org/10.1088/1367-2630/aaf9f5},
year = {2019},
month = {jan},
publisher = {IOP Publishing},
volume = {21},
number = {1},
pages = {013006},
author = {Funo, Ken and Shiraishi, Naoto and Saito, Keiji},
title = {Speed limit for open quantum systems},
journal = {New J. Phys.},
}

@article{Short2011,
doi = {10.1088/1367-2630/13/5/053009},
url = {https://doi.org/10.1088/1367-2630/13/5/053009},
year = {2011},
month = {may},
publisher = {},
volume = {13},
number = {5},
pages = {053009},
author = {Short, Anthony J},
title = {Equilibration of quantum systems and subsystems},
journal = {New J. Phys.}
}

@article{Tasaki1998,
  title = {From Quantum Dynamics to the Canonical Distribution: General Picture and a Rigorous Example},
  author = {Tasaki, Hal},
  journal = {Phys. Rev. Lett.},
  volume = {80},
  issue = {7},
  pages = {1373--1376},
  numpages = {0},
  year = {1998},
  month = {Feb},
  publisher = {American Physical Society},
  doi = {10.1103/PhysRevLett.80.1373},
  url = {https://link.aps.org/doi/10.1103/PhysRevLett.80.1373}
}

@article{Cramer2010,
doi = {10.1088/1367-2630/12/5/055020},
url = {https://doi.org/10.1088/1367-2630/12/5/055020},
year = {2010},
month = {may},
publisher = {},
volume = {12},
number = {5},
pages = {055020},
author = {Cramer, M and Eisert, J},
title = {A quantum central limit theorem for non-equilibrium systems: exact local relaxation of correlated states},
journal = {New J. Phys.}
}

@article{ShiraishiTasaki2024,
  author    = {Shiraishi, Naoto and Tasaki, Hal},
  title     = {Nature Abhors a Vacuum: A Simple Rigorous Example of Thermalization in an Isolated Macroscopic Quantum System},
  journal   = {J. Stat. Phys.},
  volume    = {191},
  pages     = {82},
  year      = {2024},
  doi       = {10.1007/s10955-024-03289-6}
}

@article{Kaufman2016,
author = {Adam M. Kaufman  and M. Eric Tai  and Alexander Lukin  and Matthew Rispoli  and Robert Schittko  and Philipp M. Preiss  and Markus Greiner },
title = {Quantum thermalization through entanglement in an isolated many-body system},
journal = {Science},
volume = {353},
number = {6301},
pages = {794-800},
year = {2016},
doi = {10.1126/science.aaf6725},
URL = {https://www.science.org/doi/abs/10.1126/science.aaf6725},
eprint = {https://www.science.org/doi/pdf/10.1126/science.aaf6725},
}

@article{Trotzky2012,
author = {Trotzky, S. and Chen, Y-A. and Flesch, A. and McCulloch, I. P. and Schollwöck, U. and Eisert, J. and Bloch, I.},
title = {Probing the relaxation towards equilibrium in an isolated strongly correlated one-dimensional Bose gas},
journal = {Nat. Phys.},
volume = {8},
number = {4},
pages = {325--330},
year = {2012},
doi = {10.1038/nphys2232},
URL = {https://doi.org/10.1038/nphys2232},
}

@misc{tasaki2000,
      title={Statistical mechanical derivation of the second law of thermodynamics}, 
      author={Hal Tasaki},
      year={2000},
      eprint={cond-mat/0009206},
      archivePrefix={arXiv},
      primaryClass={cond-mat.stat-mech},
      url={https://arxiv.org/abs/cond-mat/0009206}, 
}

@book{Gibbs1902,
  author    = {Gibbs, Josiah Willard},
  title     = {Elementary Principles in Statistical Mechanics: Developed with Special Reference to the Rational Foundation of Thermodynamics},
  year      = {1902},
  publisher = {Charles Scribner's Sons},
  address   = {New York}
}

@article{Polkovnikov2011,
title = {Microscopic diagonal entropy and its connection to basic thermodynamic relations},
journal = {Ann. Phys.},
volume = {326},
number = {2},
pages = {486-499},
year = {2011},
issn = {0003-4916},
doi = {https://doi.org/10.1016/j.aop.2010.08.004},
url = {https://www.sciencedirect.com/science/article/pii/S0003491610001557},
author = {Anatoli Polkovnikov},
}

@article{Hokkyo2025,
  title = {Universal Upper Bound on Ergotropy and No-Go Theorem by the Eigenstate Thermalization Hypothesis},
  author = {Hokkyo, Akihiro and Ueda, Masahito},
  journal = {Phys. Rev. Lett.},
  volume = {134},
  issue = {1},
  pages = {010406},
  numpages = {6},
  year = {2025},
  month = {Jan},
  publisher = {American Physical Society},
  doi = {10.1103/PhysRevLett.134.010406},
  url = {https://link.aps.org/doi/10.1103/PhysRevLett.134.010406}
}

@article{SafranekDeutschAguirre2019,
  title = {Quantum coarse-grained entropy and thermalization in closed systems},
  author = {\ifmmode \check{S}\else \v{S}\fi{}afr\'anek, Dominik and Deutsch, J. M. and Aguirre, Anthony},
  journal = {Phys. Rev. A},
  volume = {99},
  issue = {1},
  pages = {012103},
  numpages = {38},
  year = {2019},
  month = {Jan},
  publisher = {American Physical Society},
  doi = {10.1103/PhysRevA.99.012103},
  url = {https://link.aps.org/doi/10.1103/PhysRevA.99.012103}
}

@article{Brandao2015,
author = {Fernando Brandão  and Michał Horodecki  and Nelly Ng  and Jonathan Oppenheim  and Stephanie Wehner },
title = {The second laws of quantum thermodynamics},
journal = {PNAS},
volume = {112},
number = {11},
pages = {3275-3279},
year = {2015},
doi = {10.1073/pnas.1411728112},
URL = {https://www.pnas.org/doi/abs/10.1073/pnas.1411728112},
eprint = {https://www.pnas.org/doi/pdf/10.1073/pnas.1411728112}
}

@article{shiraishi2021,
  title = {Speed limit for open systems coupled to general environments},
  author = {Shiraishi, Naoto and Saito, Keiji},
  journal = {Phys. Rev. Res.},
  volume = {3},
  issue = {2},
  pages = {023074},
  numpages = {10},
  year = {2021},
  month = {Apr},
  publisher = {American Physical Society},
  doi = {10.1103/PhysRevResearch.3.023074},
  url = {https://link.aps.org/doi/10.1103/PhysRevResearch.3.023074}
}

@article{Tasaki2016,
  title = {Quantum Statistical Mechanical Derivation of the Second Law of Thermodynamics: A Hybrid Setting Approach},
  author = {Tasaki, Hal},
  journal = {Phys. Rev. Lett.},
  volume = {116},
  issue = {17},
  pages = {170402},
  numpages = {5},
  year = {2016},
  month = {Apr},
  publisher = {American Physical Society},
  doi = {10.1103/PhysRevLett.116.170402},
  url = {https://link.aps.org/doi/10.1103/PhysRevLett.116.170402}
}

@article{Meier2025,
  title = {Emergence of a Second Law of Thermodynamics in Isolated Quantum Systems},
  author = {Meier, Florian and Rivlin, Tom and Debarba, Tiago and Xuereb, Jake and Huber, Marcus and Lock, Maximilian P.E.},
  journal = {PRX Quantum},
  volume = {6},
  issue = {1},
  pages = {010309},
  numpages = {20},
  year = {2025},
  month = {Jan},
  publisher = {American Physical Society},
  doi = {10.1103/PRXQuantum.6.010309},
  url = {https://link.aps.org/doi/10.1103/PRXQuantum.6.010309}
}

@article{Ikeda2015,
title = {The second law of thermodynamics under unitary evolution and external operations},
journal = {Ann. Phys.},
volume = {354},
pages = {338-352},
year = {2015},
issn = {0003-4916},
doi = {https://doi.org/10.1016/j.aop.2015.01.003},
url = {https://www.sciencedirect.com/science/article/pii/S0003491615000068},
author = {Tatsuhiko N. Ikeda and Naoyuki Sakumichi and Anatoli Polkovnikov and Masahito Ueda},
}

@article{Dunkel2014,
  author  = {Dunkel, J{\"o}rn and Hilbert, Stefan},
  title   = {Consistent thermostatistics forbids negative absolute temperatures},
  journal = {Nat. Phys.},
  volume  = {10},
  pages   = {67--72},
  year    = {2014}
}

@article{SwendsenWang2015,
  author  = {Swendsen, Robert H. and Wang, Jian-Sheng},
  title   = {Gibbs volume entropy is incorrect},
  journal = {Phys. Rev. E},
  volume  = {92},
  pages   = {020103},
  year    = {2015},
  doi     = {10.1103/PhysRevE.92.020103}
}

@article{Braun2013,
  author  = {Braun, Stefan and Ronzheimer, Johannes P. and Schreiber, Michael and Hodgman, Sean S. and Rom, Takuya and Bloch, Immanuel and Schneider, Ulrich},
  title   = {Negative Absolute Temperature for Motional Degrees of Freedom},
  journal = {Science},
  volume  = {339},
  pages   = {52--55},
  year    = {2013}
}

@article{Ramsey1956,
  author  = {Ramsey, N. F.},
  title   = {Thermodynamics and Statistical Mechanics at Negative Absolute Temperatures},
  journal = {Phys. Rev.},
  volume  = {103},
  pages   = {20--28},
  year    = {1956}
}

@article{Lindblad1976,
  author       = {Lindblad, G{\"o}ran},
  title        = {On the generators of quantum dynamical semigroups},
  journal      = {Commun. Math. Phys.},
  volume       = {48},
  number       = {2},
  pages        = {119--130},
  year         = {1976},
  doi          = {10.1007/BF01608499},
}

@article{Gorini1976,
  author       = {Gorini, Vittorio and Kossakowski, Andrzej and Sudarshan, E. C. G.},
  title        = {Completely positive dynamical semigroups of N‐level systems},
  journal      = {J. Math. Phys.},
  volume       = {17},
  number       = {5},
  pages        = {821--825},
  year         = {1976},
  doi          = {10.1063/1.522979},
}

@article{Demirplak2003,
  author    = {Demirplak, Mustafa and Rice, Stuart A.},
  title     = {Adiabatic Population Transfer with Control Fields},
  journal   = {J. Phys. Chem. A},
  year      = {2003},
  volume    = {107},
  number    = {46},
  pages     = {9937--9945},
  doi       = {10.1021/jp030708a},
  publisher = {American Chemical Society}
}

@article{Berry2009,
doi = {10.1088/1751-8113/42/36/365303},
url = {https://doi.org/10.1088/1751-8113/42/36/365303},
year = {2009},
month = {aug},
publisher = {},
volume = {42},
number = {36},
pages = {365303},
author = {Berry, M V},
title = {Transitionless quantum driving},
journal = {J. Phys. A: Math. Theor.},
}

@article{Funoetal2017,
  title = {Universal Work Fluctuations During Shortcuts to Adiabaticity by Counterdiabatic Driving},
  author = {Funo, Ken and Zhang, Jing-Ning and Chatou, Cyril and Kim, Kihwan and Ueda, Masahito and del Campo, Adolfo},
  journal = {Phys. Rev. Lett.},
  volume = {118},
  issue = {10},
  pages = {100602},
  numpages = {6},
  year = {2017},
  month = {Mar},
  publisher = {American Physical Society},
  doi = {10.1103/PhysRevLett.118.100602},
  url = {https://link.aps.org/doi/10.1103/PhysRevLett.118.100602}
}

@article{Campbell2017,
  title = {Trade-Off Between Speed and Cost in Shortcuts to Adiabaticity},
  author = {Campbell, Steve and Deffner, Sebastian},
  journal = {Phys. Rev. Lett.},
  volume = {118},
  issue = {10},
  pages = {100601},
  numpages = {7},
  year = {2017},
  month = {Mar},
  publisher = {American Physical Society},
  doi = {10.1103/PhysRevLett.118.100601},
  url = {https://link.aps.org/doi/10.1103/PhysRevLett.118.100601}
}

@article{SunZhe2016,
  title = {Finite-time Landau-Zener processes and counterdiabatic driving in open systems: Beyond Born, Markov, and rotating-wave approximations},
  author = {Sun, Zhe and Zhou, Longwen and Xiao, Gaoyang and Poletti, Dario and Gong, Jiangbin},
  journal = {Phys. Rev. A},
  volume = {93},
  issue = {1},
  pages = {012121},
  numpages = {10},
  year = {2016},
  month = {Jan},
  publisher = {American Physical Society},
  doi = {10.1103/PhysRevA.93.012121},
  url = {https://link.aps.org/doi/10.1103/PhysRevA.93.012121}
}

@book{CoverThomas2006,
  author    = {Cover, Thomas M. and Thomas, Joy A.},
  title     = {Elements of Information Theory},
  edition   = {2},
  publisher = {Wiley-Interscience},
  address   = {Hoboken, NJ},
  year      = {2006},
  month     = jul,
  series    = {Wiley Series in Telecommunications and Signal Processing},
  isbn      = {978-0-471-24195-9},
  doi       = {10.1002/047174882X}
}

@book{NielsenChuang2010,
  author    = {Nielsen, Michael A. and Chuang, Isaac L.},
  title     = {Quantum Computation and Quantum Information},
  edition   = {10th Anniversary Edition},
  publisher = {Cambridge University Press},
  address   = {Cambridge, UK},
  year      = {2010},
  isbn      = {978-1-107-00217-3}
}

@article{Audenaert2007,
doi = {10.1088/1751-8113/40/28/S18},
url = {https://doi.org/10.1088/1751-8113/40/28/S18},
year = {2007},
month = {jun},
publisher = {},
volume = {40},
number = {28},
pages = {8127},
author = {Audenaert, Koenraad M R},
title = {A sharp continuity estimate for the von Neumann entropy},
journal = {J. Phys. A: Math. Theor.},
}

@book{Marshall2011,
  author    = {Marshall, Albert W. and Olkin, Ingram and Arnold, Barry C.},
  title     = {Inequalities: Theory of {Majorization} and Its Applications},
  edition   = {2},
  year      = {2011},
  publisher = {Springer},
  address   = {New York, NY},
  series    = {Springer Series in Statistics},
  doi       = {10.1007/978-0-387-68276-1},
  isbn      = {978-0-387-40087-7}
}

@book{Bhatia1997,
  author    = {Bhatia, Rajendra},
  title     = {Matrix Analysis},
  publisher = {Springer},
  address   = {New York},
  year      = {1997},
  isbn      = {978-0-387-94846-5},
  doi       = {10.1007/978-1-4612-0653-8}
}

\clearpage
\appendix
\onecolumngrid

{
\section{Thermodynamic coarse-grainings}
\label{app:thermodynamic-coarse-grainings}

In this Appendix, we provide detailed definitions and physical interpretations
of the thermodynamic coarse-grainings introduced in the main text. We consider
the system--bath, local-energy, and global-energy coarse-grainings. We also
derive the relation between the system--bath observational entropy and the
conventional entropy production of an open quantum system.

\subsection{System--bath coarse-graining}
\label{SigmaSobs}

Consider a small system $S$ interacting with a large bath $B$. The total
Hamiltonian is
\begin{align}
    H(\lambda)
    =
    H_S(\lambda)\otimes I_B
    +
    I_S\otimes H_B
    +
    H_I,
\end{align}
where $H_I$ denotes the interaction Hamiltonian. The projector
onto the $j$-th bath-energy mesoscopic window and its volume are defined by
\begin{align}
    \Pi_j^B
    &:=
    \mathbf 1_j(H_B), \qquad
    \Omega_j^B :=
    \tr_B[\Pi_j^B].
\end{align}

For a total state $\rho$, let
\begin{align}
    \rho_S
    :=
    \tr_B[\rho]
    =
    \sum_i \nu_i \hat P_i^{(\rho_S)}
\end{align}
be the spectral decomposition of the reduced system state. The projector
onto the kernel of $\rho_S$ is included in
$\{\hat P_i^{(\rho_S)}\}_i$ when necessary. We define the state-dependent
coarse-graining
\begin{align}
    \mathcal C_{\rho,H_B}
    :=
    \left\{
        \hat P_i^{(\rho_S)}\otimes\Pi_j^B
    \right\}_{ij}.
\end{align}

The corresponding probabilities and subspace dimensions are
\begin{align}
    p_{ij}
    &:=
    \tr\left[
        \left(
            \hat P_i^{(\rho_S)}\otimes\Pi_j^B
        \right)\rho
    \right],
    \\
    d_i
    &:=
    \tr_S[\hat P_i^{(\rho_S)}].
\end{align}
The associated observational entropy is
\begin{align}
\label{def:sb-obs}
    S_{\mathrm{SB}}(\rho)
    &:=
    S_{\mathrm{obs}}^{\mathcal C_{\rho,H_B}}(\rho)
    \nonumber =
    -\sum_{ij}
    p_{ij}
    \ln
    \frac{p_{ij}}{d_i\Omega_j^B}.
\end{align}

Introducing the marginal probabilities
\begin{align}
    p_i^S
    :=
    \sum_j p_{ij},
    \qquad
    p_j^B
    :=
    \sum_i p_{ij},
\end{align}
and the classical mutual information
\begin{align}
    I_{\mathrm{cl}}^{\mathcal C_{\rho,H_B}}(\rho)
    :=
    \sum_{ij}
    p_{ij}
    \ln
    \frac{p_{ij}}{p_i^S p_j^B},
\end{align}
the system--bath observational entropy can be decomposed as
\begin{align}
\label{eq:SB-decomposition}
    S_{\mathrm{SB}}(\rho)
    =
    S_{\mathrm{vN}}(\rho_S)
    +
    S_{\mathrm{obs}}^{\mathcal C_{H_B}}(\rho_B)
    -
    I_{\mathrm{cl}}^{\mathcal C_{\rho,H_B}}(\rho),
\end{align}
where
\begin{align}
    \rho_B
    :=
    \tr_S[\rho],
\end{align}
and
\begin{align}
    S_{\mathrm{obs}}^{\mathcal C_{H_B}}(\rho_B)
    :=
    -\sum_j
    p_j^B
    \ln
    \frac{p_j^B}{\Omega_j^B}.
\end{align}

We now relate this entropy to the conventional entropy production of an open
quantum system. We assume the initial product state
\begin{align}
\label{eq:SB-initial-state}
    \rho(0)
    =
    \rho_S(0)\otimes\pi_B,
    \qquad
    \pi_B
    =
    \frac{e^{-\beta\widetilde H_B}}{\widetilde Z_B},
\end{align}
where
\begin{align}
    \widetilde H_B
    :=
    \sum_j E_j\Pi_j^B, \qquad E_j :=-\beta^{-1}\ln\tr_B\!\left[\Pi_j^B e^{-\beta H_B}\right] +\beta^{-1}\ln\Omega_j^B,
\end{align}
The initial state in Eq.~\eqref{eq:SB-initial-state} describes the standard
situation in which the small system is prepared independently of a thermal
bath. It naturally arises when the system--bath interaction is initially
absent or sufficiently weak that initial correlations can be neglected. Thus, $\pi_B$ is uniform within each bath-energy window. We define the heat
absorbed by the bath by
\begin{align}
\label{def:bath-q}
    Q_\tau
    :=
    \tr_B
    \left[
        \widetilde H_B
        \left\{
            \rho_B(\tau)-\pi_B
        \right\}
    \right],
    \qquad
    \rho_B(\tau)
    :=
    \tr_S[\rho(\tau)].
\end{align}
The conventional open-system entropy production is
\begin{align}
\label{def:open-ent}
    \Sigma_\tau := D\!\left(
        \rho(\tau)
        \,\middle\|\,
        \rho_S(\tau)\otimes\pi_B
    \right).
\end{align}

Using $\ln\pi_B=-\beta\tilde H_B-\ln\tilde Z_B$, the initial product state
$\rho(0)=\rho_S(0)\otimes\pi_B$, and the unitary invariance of the von Neumann entropy, we obtain
\begin{align}
D\!\left(\rho(\tau)\middle\|\rho_S(\tau)\otimes\pi_B\right)
&=
-S_{\mathrm{vN}}(\rho(\tau))
+S_{\mathrm{vN}}(\rho_S(\tau))
-\tr[\rho_B(\tau)\ln\pi_B]
\nonumber\\
&=
S_{\mathrm{vN}}(\rho_S(\tau))
-S_{\mathrm{vN}}(\rho_S(0))
+\beta\tr\!\left[\tilde H_B\{\rho_B(\tau)-\pi_B\}\right]
\nonumber\\
&=
\Delta S_{\mathrm{vN}}(\rho_S)+\beta Q_\tau .
\end{align}

The change in the system--bath observational entropy is denoted by
\begin{align}
    \Delta S_{\mathrm{SB}}(\tau)
    &:=
    S_{\mathrm{obs}}^{\mathcal C_{\rho(\tau),H_B}}
    (\rho(\tau))
    -
    S_{\mathrm{obs}}^{\mathcal C_{\rho(0),H_B}}
    (\rho(0)).
\end{align}
Since the initial state is a product state, its classical mutual information
vanishes. Equation~\eqref{eq:SB-decomposition} therefore gives
\begin{align}
    &
    \Delta S_{\mathrm{SB}}(\tau)
    +
    I_{\mathrm{cl}}^{\mathcal C_{\rho(\tau),H_B}}
    (\rho(\tau))
    \nonumber\\
    &=
    S_{\mathrm{vN}}(\rho_S(\tau))
    -
    S_{\mathrm{vN}}(\rho_S(0))
    +
    S_{\mathrm{obs}}^{\mathcal C_{H_B}}(\rho_B(\tau))
    -
    S_{\mathrm{obs}}^{\mathcal C_{H_B}}(\pi_B).
\end{align}
Consequently,
\begin{align}
\label{eq:Sigma-minus-SB}
    &
    \Sigma_\tau
    -
    \Delta S_{\mathrm{SB}}(\tau)
    -
    I_{\mathrm{cl}}^{\mathcal C_{\rho(\tau),H_B}}
    (\rho(\tau))
    \nonumber\\
    &=
    \beta Q_\tau
    -
    S_{\mathrm{obs}}^{\mathcal C_{H_B}}(\rho_B(\tau))
    +
    S_{\mathrm{obs}}^{\mathcal C_{H_B}}(\pi_B).
\end{align}

Define the final and initial bath-energy distributions by
\begin{align}
    \alpha_j
    &:=
    \tr_B[\Pi_j^B\rho_B(\tau)],
    \\
    \pi_j
    &:=
    \tr_B[\Pi_j^B\pi_B]
    =
    \frac{
        \Omega_j^B e^{-\beta E_j}
    }{
        \widetilde Z_B
    }.
\end{align}
Using
\begin{align}
    \ln\frac{\pi_j}{\Omega_j^B}
    =
    -\beta E_j-\ln\widetilde Z_B,
\end{align}
the right-hand side of Eq.~\eqref{eq:Sigma-minus-SB} becomes
\begin{align}
    &
    \beta Q_\tau
    -
    S_{\mathrm{obs}}^{\mathcal C_{H_B}}(\rho_B(\tau))
    +
    S_{\mathrm{obs}}^{\mathcal C_{H_B}}(\pi_B)
    \nonumber\\
    &=
    \sum_j
    \alpha_j
    \ln
    \frac{\alpha_j}{\pi_j}
    =
    D(\alpha\Vert\pi).
\end{align}
We therefore obtain the exact identity
\begin{align}
\label{eq:Sigma-SB-identity}
    \Sigma_\tau
    =
    \Delta S_{\mathrm{SB}}(\tau)
    +
    I_{\mathrm{cl}}^{\mathcal C_{\rho(\tau),H_B}}
    (\rho(\tau))
    +
    D(\alpha\Vert\pi).
\end{align}
Since both the classical mutual information and the relative entropy are
nonnegative, this identity immediately yields
\begin{align}
\label{SigmaSobseq}
    \Sigma_\tau
    &\geq
    \Delta S_{\mathrm{SB}}(\tau)
    +
    I_{\mathrm{cl}}^{\mathcal C_{\rho(\tau),H_B}}
    (\rho(\tau))
    \nonumber\\
    &\geq
    \Delta S_{\mathrm{SB}}(\tau).
\end{align}
Thus, the system--bath observational entropy provides a coarse-grained
description of the conventional entropy production. If the final bath-energy
distribution remains close to its initial equilibrium distribution, then
$D(\alpha\Vert\pi)$ is small and Eq.~\eqref{eq:Sigma-SB-identity} gives
\begin{align}
    \Sigma_\tau
    \simeq
    \Delta S_{\mathrm{SB}}(\tau)
    +
    I_{\mathrm{cl}}^{\mathcal C_{\rho(\tau),H_B}}
    (\rho(\tau)).
\end{align}

\subsection{Local-energy coarse-graining}

We next consider an isolated many-body system divided into $m$ spatial
regions,
\begin{align}
    \mathcal H
    =
    \bigotimes_{\ell=1}^{m}\mathcal H_\ell.
\end{align}
The Hamiltonian is decomposed as
\begin{align}
    H
    =
    \sum_{\ell=1}^{m}H_\ell
    +
    H_{\mathrm{int}},
\end{align}
where $H_\ell$ is the local Hamiltonian of region $\ell$, and
$H_{\mathrm{int}}$ describes interactions between different regions. Let
\begin{align}
    H_\ell
    =
    \sum_{E_\ell}
    E_\ell\hat P_{E_\ell}^{(\ell)}
\end{align}
be the spectral decomposition of $H_\ell$. The projectors resolving the
local-energy profile are
\begin{align}
    \hat P_{\vec E}
    :=
    \bigotimes_{\ell=1}^{m}
    \hat P_{E_\ell}^{(\ell)},
    \qquad
    \vec E
    :=
    (E_1,\ldots,E_m).
\end{align}
The corresponding coarse-graining is
\begin{align}
    \mathcal C_{\mathrm{loc}H}
    :=
    \left\{
        \hat P_{\vec E}
    \right\}_{\vec E}.
\end{align}

The factorized observational entropy is defined by
\begin{align}
\label{eq:FOE_def}
    S_{\mathrm{FOE}}(\rho)
    &:=
    S_{\mathrm{obs}}^{\mathcal C_{\mathrm{loc}H}}(\rho) = 
    -\sum_{\vec E}
    p_{\vec E}
    \ln
    \frac{p_{\vec E}}{V_{\vec E}},
\end{align}
where
\begin{align}
    p_{\vec E}
    &:=
    \tr[\hat P_{\vec E}\rho],
    \\
    V_{\vec E}
    &:=
    \tr[\hat P_{\vec E}]
    =
    \prod_{\ell=1}^{m}
    \tr[\hat P_{E_\ell}^{(\ell)}].
\end{align}
If the state is supported within a single local-energy macrospace
$\vec E$, then
\begin{align}
    S_{\mathrm{FOE}}(\rho)
    =
    \ln V_{\vec E}
    =
    \sum_{\ell=1}^{m}
    \ln
    \tr[\hat P_{E_\ell}^{(\ell)}].
\end{align}
Thus, $S_{\mathrm{FOE}}$ reduces to the sum of the microcanonical entropies
of the individual regions. More generally, it quantifies the entropy
associated with a spatially inhomogeneous energy profile.

An initial macrostate with respect to
$\mathcal C_{\mathrm{loc}H}$ has the form
\begin{align}
\label{eq:FOE-initial-state}
    \rho(0)
    =
    \bigotimes_{\ell=1}^{m}\rho_\ell,
    \qquad
    \rho_\ell
    =
    \sum_{E_\ell}
    p_{\ell,E_\ell}
    \frac{
        \hat P_{E_\ell}^{(\ell)}
    }{
        \tr[\hat P_{E_\ell}^{(\ell)}]
    }.
\end{align}
For nondegenerate local Hamiltonians, this condition reduces to
\begin{align}
    [\rho_\ell,H_\ell]=0.
\end{align}
Such states naturally describe local equilibrium: each spatial region has
relaxed with respect to its own local Hamiltonian, while the system as a
whole remains out of equilibrium. This description is appropriate when
\begin{align}
    \tau_{\mathrm{local}}
    \ll
    \tau_{\mathrm{transport}},
\end{align}
so that local equilibration occurs much faster than transport between
different regions. It therefore provides the natural coarse-grained
description of hydrodynamic relaxation and of systems prepared by joining
subsystems initially held at different temperatures.

\subsection{Global-energy coarse-graining and diagonal entropy}

Finally, consider the spectral decomposition of the global Hamiltonian,
\begin{align}
    H(\lambda)
    =
    \sum_i
    E_i(\lambda)
    \hat P_i^{(E)}(\lambda).
\end{align}
The global-energy coarse-graining is
\begin{align}
    \mathcal C_{H(\lambda)}
    :=
    \left\{
        \hat P_i^{(E)}(\lambda)
    \right\}_i.
\end{align}
The corresponding observational entropy is
\begin{align}
    S_D(\rho;H(\lambda))
    &:=
    S_{\mathrm{obs}}^{\mathcal C_{H(\lambda)}}(\rho) = 
    -\sum_i
    p_i
    \ln
    \frac{p_i}{V_i(\lambda)},
\end{align}
where
\begin{align}
    p_i
    &:=
    \tr[
        \hat P_i^{(E)}(\lambda)\rho
    ],
    \\
    V_i(\lambda)
    &:=
    \tr[
        \hat P_i^{(E)}(\lambda)
    ].
\end{align}
For a nondegenerate Hamiltonian, $V_i(\lambda)=1$, and this expression
reduces to the conventional diagonal entropy
\begin{align}
    S_d(\rho)
    =
    -\sum_i
    \rho_{ii}
    \ln\rho_{ii}.
\end{align}

The global-energy coarse-graining is natural for driven isolated systems,
because it retains only the occupation probabilities of the instantaneous
energy eigenspaces. Energy-diagonal states include equilibrium Gibbs and
microcanonical states. They also arise as diagonal ensembles describing
long-time averages after quantum quenches. The diagonal entropy therefore
provides a coarse-grained entropy suitable for both equilibrium states and
effective stationary states generated by isolated unitary dynamics.
}

\section{A unitary transformation that preserves the Gibbs entropy}
\label{app-GibbsUni}

For any passive state $\rho$, we show that the unitary operator that preserves the Gibbs entropy is unique up to phases and is given by
\begin{align}
U_{\mathrm{ad}} := \sum_i \ketbra{i_\tau}{i_0}.
\end{align}
Here $\{\ket{i_0}\}_{i=1}^{D}$ and $\{\ket{i_{\tau}}\}_{i=1}^{D}$ are the eigenstates of the initial Hamiltonian $H(\lambda_0)$ and the final Hamiltonian $H(\lambda_\tau)$, respectively, ordered by increasing eigenvalues.\\

Let $U$ be a unitary operator that preserves the Gibbs entropy. Then, for any passive state $\rho$,
\begin{align}
\sum_{i=1}^d \bra{i_{\tau}}U\rho U^{\dagger}\ket{i_{\tau}} \ln i
= \sum_{i=1}^d \bra{i_{0}}\rho \ket{i_{0}} \ln i .
\end{align}
It follows that, for the extended microcanonical state $\rho_{em}(\Omega) := \frac{1}{\Omega}\sum_{i=1}^\Omega \ketbra{i_0}$ $(1\le \Omega \le d)$,
\begin{align}
\sum_{i=1}^d \sum_{j=1}^{\Omega} \left|\bra{i_{\tau}}U\ket{j_0}\right|^2 \ln i
= \sum_{i=1}^{\Omega} \ln i .
\end{align}
Starting with the case $\Omega = 1$, the above equation reduces to
\begin{align}
\sum_{i=1}^d \left|\bra{i_{\tau}}U\ket{1_0}\right|^2 \ln i = 0 .
\end{align}
If $U$ satisfied $\bra{i_{\tau}}U\ket{1_0} \neq 0$ for some $i>1$, then 
\begin{align}
\sum_{i=1}^d \left|\bra{i_{\tau}}U\ket{1_0}\right|^2 \ln i > 0
\end{align}
would hold, which leads to a contradiction. Therefore, $\bra{i_{\tau}}U\ket{1_0} = 0$ for $i>1$ and $|\bra{1_{\tau}}U\ket{1_0}| = 1$. By unitarity, $\sum_i \left|\bra{1_{\tau}} U \ket{i_0}\right|^2 = 1$, and hence $|\bra{1_{\tau}} U \ket{i_0}|^2 = \delta_{1i}$.
Next, considering the case $\Omega = 2$, we obtain
\begin{align}
\sum_{i=2}^d \left|\bra{i_{\tau}}U\ket{2_0}\right|^2 \ln i = \ln 2 .
\end{align}
By the same reasoning as in the case $\Omega = 1$, we obtain $\left|\bra{i_{\tau}}U\ket{2_0}\right| = \left|\bra{2_{\tau}}U\ket{i_0}\right| = \delta_{i2}$. Repeating the same argument sequentially up to $\Omega = d-1$, we obtain $\left|\bra{i_{\tau}}U\ket{j_0}\right|=\delta_{ij}$.

\section{The Proof of Eq.~\eqref{AppA}}
\label{app-Wq}

The derivation follows Ref.~\cite{VuSaito2023}. At first, we show that $\mathcal{W}_q(\rho, \sigma) \le  \f{1}{2} \sum_i |\lambda_{\rho}^{\downarrow}(i) - \lambda^{\downarrow}_{\sigma}(i)|$. We consider 
\begin{align}
    \rho &= \sum_i \lambda_{\rho}(i) \ketbra{i_{\rho}},\quad \sigma = \sum_i \lambda_{\sigma}(i) \ketbra{i_{\sigma}}.
\end{align}
Here $\lambda_{X}(i)$ denotes the $i$th smallest eigenvalue of the operator $X$, and $\ket{i_X}$ is the corresponding eigenvector. we consider $W = \sum_i \ketbra{i_{\rho}}{i_{\sigma}}$. By definition, we obtain
\begin{align}
    \mathcal{W}_q(\rho, \sigma) = \f{1}{2}\min_{VV^{\dagger}=1} \|V\rho V^{\dagger} - \sigma\| \le \f{1}{2} \|W\rho W^{\dagger} - \sigma\| = \f{1}{2} \sum_i |\lambda_{\rho}^{\downarrow}(i) - \lambda^{\downarrow}_{\sigma}(i)|.
\end{align}
Next, we show $\mathcal{W}_q(\rho, \sigma) \ge  \f{1}{2} \sum_i |\lambda_{\rho}^{\downarrow}(i) - \lambda^{\downarrow}_{\sigma}(i)|$. For any Hermitian operators $A$ and $B$, let $\eta_i(A)$ and $\eta_i(B)$ denote their singular values arranged in nondecreasing order. Then
\begin{align}
    \|A-B\| = \sum_i |\eta_i(A-B)| \ge \sum_i |\eta_i(A) - \eta_i(B)|.
\end{align}
holds\cite{Bhatia1997}. By setting $A=V\rho V^{\dagger}$,$B=\sigma$, we obtain the desired inequality. Therefore, $\mathcal{W}_q(\rho, \sigma) =  \f{1}{2} \sum_i |\lambda_{\rho}^{\downarrow}(i) - \lambda^{\downarrow}_{\sigma}(i)|$ follows.

{
\section{Bounds for non-thermodynamic entropy production}
\label{Non-thermo}

Here, as an application of the main results in Eqs.~\eqref{general-res} and \eqref{general-res2} to a non-thermodynamic entropy, we consider the entanglement entropy.

Let \(\ket{\Psi}_{AB}\) be a pure state on the bipartite Hilbert space \(\mathcal{H}_A \otimes \mathcal{H}_B\). We write its Schmidt decomposition as
\begin{align}
    \ket{\Psi}_{AB}
    =
    \sum_{\alpha}\sum_{a=1}^{d_\alpha}
    \sqrt{\lambda_\alpha}\,
    \ket{\alpha,a}_A \otimes \ket{\alpha,a}_B,
    \qquad
    \sum_{\alpha} d_\alpha \lambda_\alpha = 1 .
\end{align}
Here, \(\lambda_\alpha\) denotes a distinct Schmidt eigenvalue, and \(d_\alpha\) is its degeneracy. The reduced density matrix on subsystem \(A\) is then given by
\begin{align}
    \rho_A
    =
    \tr_B[\ket{\Psi}\bra{\Psi}]
    =
    \sum_{\alpha}\lambda_\alpha \hat{P}_\alpha^A,
    \qquad
    \hat{P}_\alpha^A
    =
    \sum_{a=1}^{d_\alpha}
    \ket{\alpha,a}_A\bra{\alpha,a}_A .
\end{align}

We define projectors on the total Hilbert space \(\mathcal{H}_A\otimes\mathcal{H}_B\) by
\begin{align}
    \hat{Q}_\alpha^{AB}
    :=
    \sum_{a=1}^{d_\alpha}
    \left(
        \ket{\alpha,a}_A \otimes \ket{\alpha,a}_B
    \right)
    \left(
        {}_A\bra{\alpha,a} \otimes {}_B\bra{\alpha,a}
    \right),
\end{align}
and
\begin{align}
    \hat{Q}_0^{AB}
    :=
    \hat{1}_{AB}
    -
    \sum_\alpha \hat{Q}_\alpha^{AB}.
\end{align}
These projectors define a coarse-graining of the total Hilbert space. We denote this state-dependent coarse-graining by
\begin{align}
    \mathcal{C}_{\mathrm{ent}}(\Psi)
    :=
    \{\hat{Q}_\alpha^{AB}\}_{\alpha}
    \cup
    \{\hat{Q}_0^{AB}\}.
\end{align}

For the pure state \(\rho_{AB}=\ket{\Psi}\bra{\Psi}\), the probability and volume associated with \(\hat{Q}_\alpha^{AB}\) are respectively
\begin{align}
    p_\alpha
    =
    \tr[\hat{Q}_\alpha^{AB}\rho_{AB}]
    =
    d_\alpha\lambda_\alpha,
    \qquad
    V_\alpha
    =
    \tr[\hat{Q}_\alpha^{AB}]
    =
    d_\alpha .
\end{align}
Moreover, \(p_0=0\). Therefore, the observational entropy of the total system with respect to the coarse-graining \(\mathcal{C}_{\mathrm{ent}}(\Psi)\) is
\begin{align}
    S_{\mathrm{obs}}^{\mathcal{C}_{\mathrm{ent}}(\Psi)}(\rho_{AB})
    &=
    -
    \sum_\alpha
    p_\alpha
    \ln\frac{p_\alpha}{V_\alpha} \\
    &=
    -
    \sum_\alpha
    d_\alpha\lambda_\alpha
    \ln\lambda_\alpha \\
    &=
    S_{\mathrm{vN}}(\rho_A).
\end{align}
Since \(\rho_{AB}\) is pure, \(S_{\mathrm{vN}}(\rho_A)\) is the entanglement entropy between \(A\) and \(B\). Hence,
\begin{align}
    S_{\mathrm{obs}}^{\mathcal{C}_{\mathrm{ent}}(\Psi)}
    (\ket{\Psi}\bra{\Psi})
    =
    S_{\mathrm{ent}}^{(A:B)}(\ket{\Psi}).
\end{align}
Thus, the entanglement entropy of a bipartite pure state can be regarded as the observational entropy of the total system associated with the Schmidt-correlated coarse-graining \(\mathcal{C}_{\mathrm{ent}}(\Psi)\).

Using this observation, the main results in Eqs.~\eqref{general-res} and \eqref{general-res2} yield bounds on the entanglement entropy production
\begin{align}
    \Delta S_{\mathrm{ent}}^{(A:B)}(\tau)
    :=
    S_{\mathrm{ent}}^{(A:B)}(\ket{\Psi(\tau)})
    -
    S_{\mathrm{ent}}^{(A:B)}(\ket{\Psi(0)}).
\end{align}
Specifically, assuming that the initial state is a product state,
\begin{align}
    \ket{\Psi(0)}
    =
    \ket{\psi}_A \otimes \ket{\psi'}_B,
\end{align}
we obtain
\begin{align}
    \Delta S_{\mathrm{ent}}^{(A:B)}(\tau) = S_{\mathrm{ent}}^{(A:B)}(\ket{\Psi(\tau)})
    \ge
    2
    \mathcal{W}_q
    \left(
        \rho_{\mathcal{C}_{\mathrm{ent}}(\Psi(\tau))}(\tau),
        \ket{\Psi(0)}\bra{\Psi(0)}
    \right)^2 ,
\end{align}
and
\begin{align}
    \tau
    \ge
    {\frac{
        2
        \mathcal{W}_q
        \left(
            \rho_{\mathcal{C}_{\mathrm{ent}}(\Psi(\tau))}(\tau),
            \ket{\Psi(0)}\bra{\Psi(0)}
        \right)^2
    }{
        \bar{\sigma}_{\mathrm{ent}}^{(A:B)}
    }},
    \qquad
    \bar{\sigma}_{\mathrm{ent}}^{(A:B)}
    :=
    \frac{
        \Delta S_{\mathrm{ent}}^{(A:B)}(\tau)
    }{
        \tau
    } .
\end{align}
}

\section{The Lenard--Tasaki inequality}
\label{app-q-p}
The derivation follows Ref. \cite{tasaki2000}.
Consider an observable $F$ and a probability distribution $u$ satisfying $i<j \Rightarrow F_i \le F_j$ and $u(i)\ge u(j)$. Let $T$ be a doubly stochastic matrix and define $v=Tu$. We show that $\langle F\rangle_v \ge \langle F\rangle_u$. Introducing $u_a(i)=\Theta(a-i)/a$, we expand $u = \sum_a c_a u_a, c_a := a\{u(a) - u(a+1)\}, v_a := T u_a$. Then, we obtain 
\begin{align}
    \langle F \rangle_v - \langle F \rangle_u &= \sum_i F_i (v(i) - u(i))\nonumber \\
    &= \sum_a c_a \sum_i F_i (v_a(i) - u_a(i))\nonumber \\
    &= \sum_a c_a \left[\sum_{i \ge a+1} F_i (v_a(i) - u_a(i)) + \sum_{i \le a} F_i (v_a(i) - u_a(i))\right]\nonumber \\
    &\ge \sum_a c_a \left[F_{a+1} \sum_{i \ge a+1}  (v_a(i) - u_a(i)) + F_a \sum_{i \le a} (v_a(i) - u_a(i))\right]\nonumber \\
    &\ge \sum_a c_a F_{a}\left[\sum_{i \ge a+1}  (v_a(i) - u_a(i)) + \sum_{i \le a} (v_a(i) - u_a(i))\right]\nonumber \\
    &= 0.
\end{align}
{The first inequality follows from the doubly stochasticity of the map}, which implies $i \ge a+1 \Rightarrow v_a(i) - u_a(i) \ge 0, \quad i \le a \Rightarrow v_a(i) - u_a(i) \le 0$. The transition from $p^{\downarrow}_0$ to $p^{\downarrow}_t$ can be realized by a doubly stochastic matrix. Choosing the observable as $F_i=-\ln p^{\downarrow}_t(i)$, we note that $p^{\downarrow}_t$ is ordered in descending order, so ${F_i}$ is non-decreasing in $i$. Since $p^{\downarrow}_0$ is also descending, all assumptions of the theorem are satisfied. Therefore
\begin{align}
    \sum_i -\ln p^{\downarrow}_t(i) (p^{\downarrow}_t(i) - p^{\downarrow}_0(i)) \ge 0.
\end{align}

\label{sec:app_momeqs}
\label{sec:app_NI}

\end{document}